\documentclass[aps,prl,twocolumn,superscriptaddress]{revtex4-1}

\usepackage{amsmath}
\usepackage{color,xcolor}
\usepackage{graphicx}
\usepackage[colorlinks=true,linkcolor=blue,urlcolor=blue,citecolor=blue]{hyperref}
\usepackage{bm}
\usepackage{amsfonts}
\usepackage{soul}
\usepackage{braket}
\usepackage{lipsum}
\usepackage{siunitx}
\newcommand{\m}{N}
\newcommand{\Eenv}{\mathcal{E}}
\newcommand{\Nlabel}{{}}

\begin{document}
\title{ Quantum oscillations  in the zeroth Landau Level and the serpentine Landau fan}
\author{T. Devakul}
\affiliation{Department of Physics, Massachusetts Institute of Technology, Cambridge, Massachusetts 02139, USA}
\affiliation{Department of Physics, Princeton University, Princeton, New Jersey 08540, USA}
	\author{Yves H. Kwan}
\affiliation{Rudolf Peierls Centre for Theoretical Physics,  Clarendon Laboratory, Oxford OX1 3PU, UK}
\author{S. L. Sondhi}
\affiliation{Department of Physics, Princeton University, Princeton, New Jersey 08540, USA}
		\author{S. A. Parameswaran}
\affiliation{Rudolf Peierls Centre for Theoretical Physics,  Clarendon Laboratory, Oxford OX1 3PU, UK}

\date{\today }

\begin{abstract}
We identify an unusual mechanism for quantum oscillations in nodal semimetals, driven by a {\it single} pair of Landau levels periodically closing their gap at the Fermi energy  as a magnetic field is varied. These `zero Landau level' quantum oscillations (ZQOs) appear in the nodal limit where the zero-field Fermi volume vanishes, and have distinctive  periodicity and temperature dependence.   We link the Landau spectrum of a two-dimensional (2D) nodal semimetal to the  Rabi model, and show by exact solution that across the entire Landau  fan, pairs of opposite-parity  Landau levels are  intertwined in a `serpentine' manner. 
We propose 2D surfaces of topological crystalline insulators as natural settings for ZQOs.
In certain 3D nodal semimetals, ZQOs lead to oscillations of anomaly physics.
We propose a transport measurement capable of observing such oscillations, which we demonstrate numerically.

\end{abstract}

\maketitle

\emph{Introduction.---} Quantum oscillations (QOs) --- the periodic modulation of transport~\cite{SdH} and thermodynamic~\cite{dHvA}  properties of materials as an external magnetic field $\bm{B}$  is varied--- are among the most striking manifestations of quantum mechanics in the solid state.  
The oscillation period depends on the shape of the Fermi surface (FS), while  information on FS parameters  and  scattering mechanisms  can be extracted from the oscillation amplitude and its temperature dependence~\cite{shoenberg_1984}. The utility of  QOs as a probe of electronic structure rests on two theoretical pillars. 
The first is Onsager's result~\cite{Onsager}, \begin{equation}\label{eq:Onsager}
\Delta^{\text{O}}_{1/B} = 2\pi \mathcal{S}_e^{-1},
\end{equation}
relating their periodicity in $1/B$ to an extremal cross-sectional FS area $\mathcal{S}_e$ in a plane normal to $\bm{B}$ (we set $\hbar=e=1$). The second is the  Lifshitz-Kosevich formula~\cite{LK} 
\begin{equation}\label{eq:LK}
R_{\text{LK}}(T)  =  \frac{2\pi^2T}{\omega_c\sinh ({2\pi^2 T}/{\omega_c})}
\end{equation}
with $k_B=1$, relating the temperature ($T$) dependence of the oscillation amplitude (shown here for the first harmonic) to the cyclotron frequency $\omega_c = B/m^*$, allowing the extraction of the effective mass $m^*$, averaged over $\mathcal{S}_e$.

Topological insulators and nodal semimetals~\cite{Armitage2018} modify this picture.  The effective-mass approximation implicitly assumed in~\eqref{eq:Onsager} and~\eqref{eq:LK} is violated  by Weyl or Dirac dispersions, and the QO phase is altered by electronic Berry phases of topologically nontrivial bands~\cite{Mikitik99}. Although many features may be captured by adapting the semiclassical theory~\cite{AG1,AG2}, the latter assumes the existence of a FS at $\bm{B}=0$. It does not readily apply if a FS is absent, as in an insulator, or when it  shrinks to a point in two dimensions (2D) or a  point or line in 3D,  situations which cannot be generated solely by the intersection of unhybridized bands.
Such systems enter the quantum limit for any $\bm{B}\neq0$, necessitating a full solution of the Landau level (LL) spectrum.

In 2D, our focus in much of this work, a linear crossing of a pair of energy bands at $E=0$ is generically described by the 2D Dirac equation. 
Viewed in isolation, such a dispersion gives rise to a fan of LLs whose energies evolve   $\propto \sqrt{B}$, with the exception of a single $\bm{B}$-independent LL at $E=0$.  
While $1/{B}$-periodic QOs emerge at any finite doping,  at charge neutrality --- corresponding to a point-node FS for $\bm{B}\to 0$ --- conventional QOs are absent since no LLs cross the Fermi energy at $E_F=0$. 
However, in bulk systems, such nodal points always appear in pairs that are usually separated in the Brillouin zone (BZ) for symmetry reasons. 
Consequently, in principle there is always mixing of LLs emerging from distinct nodes.
The intuitive expectation is that as $B$ is increased, the zero-energy LLs from  distinct nodes will experience increasingly strong  mutual level repulsion,  pushing them away from $E_F$ without intersecting it~\cite{balatskii1986chiral, WeylAnnihilation,ChanLee, Bednik}.

Here, we show that a class of nodal-point semimetals  violate this expectation: the zero-energy LLs oscillate about each other in energy as $B$ is varied, leading to  robust QOs as they repeatedly intersect $E_F$. The simplest instance of this general phenomenon of zero-LL QOs (ZQOs) occurs when a pair of 2D 
parabolic bands with effective mass $m$ undergo a band inversion of strength $\Delta$. This leads to a degenerate nodal ring at $|\bm{k}| = k_0 \equiv\sqrt{m\Delta}$ which, assuming inversion or mirror symmetry, is generically gapped by hybridization except at a  pair of nodal points at $\pm\bm{k}_0 =\pm k_0 \hat{\bm{x}}$ with  anisotropic velocities $v_x = v_c \equiv\sqrt{\Delta/m}, v_y=v$. We relate the associated LL problem to the Rabi model of quantum optics, and by exact solution identify a sequence of QOs at charge neutrality. Although $\mathcal{S}_e=0$, these zero-LL QOs show $1/B$-periodicity controlled by  the  FS area $\mathcal{S}_0 = \pi k_0^2$ of the unhybridized bands at $E_F=0$, corrected by a factor: 
\begin{equation}\label{eq:ZLL}
\Delta^{\text{ZLL}}_{1/B}  
={2\pi}\gamma^2 \mathcal{S}_0^{-1}, \,\,\,\, \gamma = 1/{\sqrt{1-v^2/v_c^2}}. 
\end{equation}
The ZQOs only occur if $v<v_c$, and disappear above $B_0 = \mathcal{S}_0/\pi\gamma^2$. They  originate from the `serpentine' motion of two LLs that straddle  $E_F$ and periodically open and close their gap while remaining bounded within an envelope $\Eenv(B)$ and separated from other LLs. At temperatures where these two states dominate the spectrum, ZQOs show non-Lifshitz-Kosevich (LK) behaviour
\begin{equation}
\label{eq:RZLL}
R_{\text{ZLL}}(T) = \frac{\Eenv(B)}{2T} \tanh^2 \frac{\Eenv(B)}{2T}.
\end{equation}
The explicit form of $\Eenv(B)$, given below,  cannot be modeled by a  cyclotron frequency with an effective mass, and as $B\to 0$ has the non-perturbative form $\Eenv(B)\sim e^{-B^*/B}$.

ZQOs differ from other proposed routes to  QOs at $\mathcal{S}_e=0$.  The purely orbital origin of ZQOs and the associated serpentine Landau fan distinguishes their mechanism from recent proposals of Zeeman-driven `LL inversions'~\cite{Wang2020}. As they do not invoke  surface states stemming from a topologically nontrivial band structure,  ZQOs differ from Fermi-arc QOs in Weyl/Dirac semimetal slabs~\cite{Potter2014}. The $e^{-B^*/B}$ dependence of the ZQO envelope and their link to the unhybridized FS area are reminiscent of  QOs in narrow-gap insulators~\cite{Knolle2015,Knolle2017a, Zhang2016, Pal2016, Pal2017, Ram2017,Ram2019, Grubinskas2018, Shen2018, Lu2020, Falkovsky2011, Alisultanov2016}. Both phenomena can be understood by extending the semiclassical approach to include phase-coherent tunneling through classically forbidden regions in the BZ. However, the serpentine fan, the $\gamma$-dependent period,  and the gap-closings at $E_F$ are special to ZQOs. Unlike QOs in insulators, whose effect on the conductivity  is suppressed by the activation gap, ZQOs remain observable in transport even as $T\to 0$. ZQOs thus represent a distinct class of magnetic oscillation phenomenon.

Below, we substantiate these claims by introducing and exactly solving a model that exhibits ZQOs, derive \eqref{eq:ZLL} and \eqref{eq:RZLL}, and explain their connection to the serpentine structure across the Landau fan.
We argue that key features  persist even upon relaxing simplifying assumptions of the solvable limit.  
We close with a discussion of the observability and interpretation of ZQOs.

\emph{Model and LL Spectrum.---} As noted above, the simplest model showing ZQOs begins with a $k\cdot p$ Hamiltonian for a spinless band-inverted electron in 2D,
\begin{equation}
H(\bm{k}) = \left(\frac{|\bm{k}|^2}{2m} - \frac{\Delta}{2}\right)\tau^z + v k_y \tau^y,
\label{eq:H0}
\end{equation}
where $\Delta,v>0$, and $\tau^{\alpha}$ are Pauli matrices acting in orbital space. 
At half filling and low energy, this yields the advertised Dirac cones at $\pm\bm{k}_0$,
with anisotropic velocities $(v_x, v_y) =(v_c, v)$; we present results in terms of $k_0,v,v_c$. 
The Dirac points are protected by inversion (parity) $\mathcal{P}:\tau^z\otimes (\bm{k}\rightarrow-\bm{k})$ and time reversal $\mathcal{T}: K \otimes (\bm{k}\rightarrow -\bm{k})$, where $K$ is complex conjugation.

\begin{figure}
\includegraphics[width=0.45\textwidth]{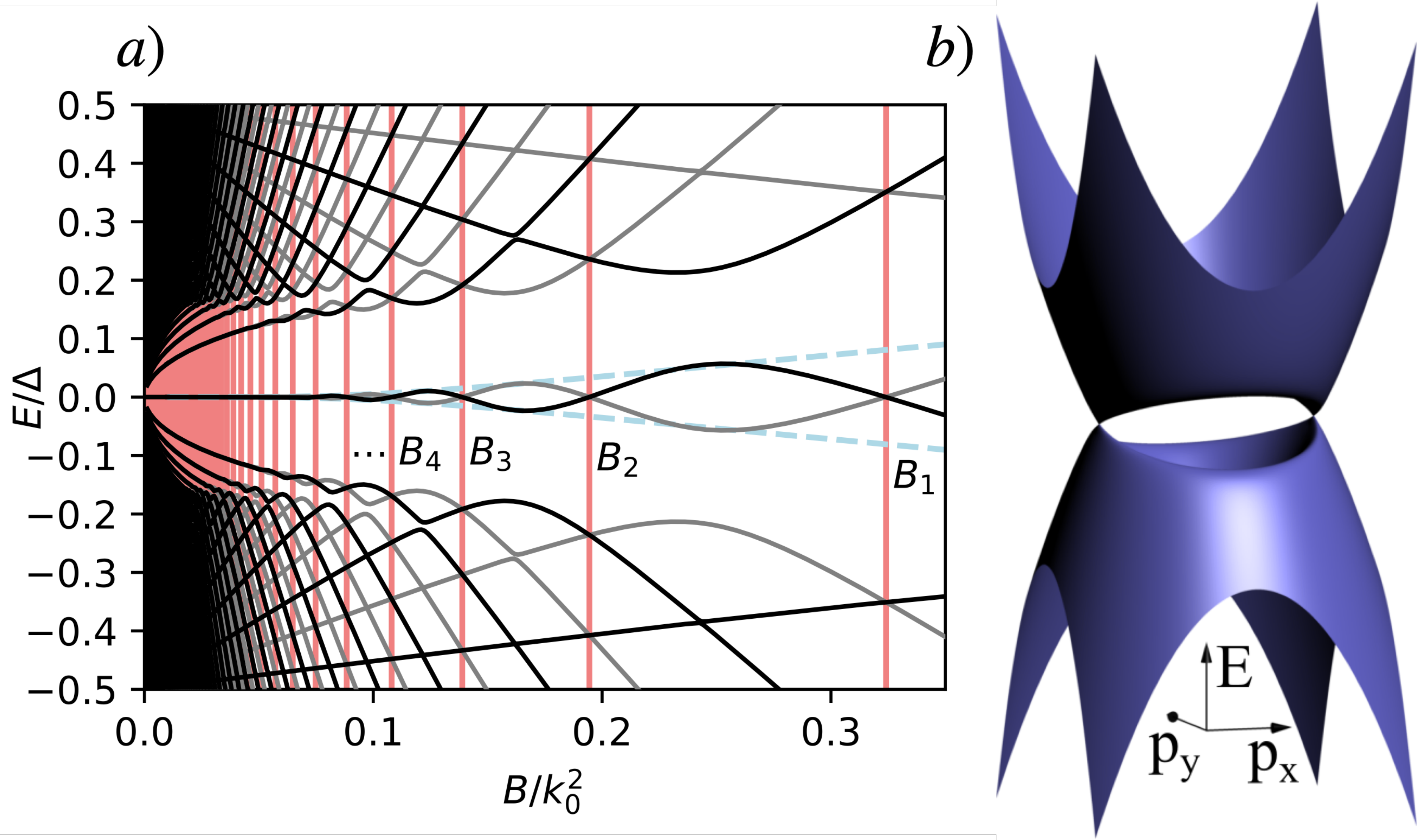}
\caption{$(a)$ Serpentine Landau level  (LL) fan computed for $v=v_c/6$. Even and odd parity LLs (black/gray lines) repeatedly self-intersect within a widening envelope as $B$ is varied. ZQOs originate from a pair that repeatedly intersects the Fermi energy ($E$=0) at magic fields $B_\m$ (red lines), where the LL problem simplifies. 
The oscillation magnitude is well approximated by $\pm\mathcal{E}(B)$ derived in the main text (dashed lines). 
$(b)$  Zero-field dispersion of  $H$ (\ref{eq:H0}).
}
\label{fig:energy}
\end{figure}

We incorporate a magnetic field $B \hat{\bm{z}}= \bm{\nabla}\times \bm{A}$ via Peierls substitution $\bm{k}\rightarrow \bm{\pi}= \bm{k}-\bm{A}$, so that $[\pi_x,\pi_y] = i B$. We first obtain the LL spectrum numerically  (Fig~\ref{fig:energy}), by truncating in the  basis defined by $a^\dagger a\ket{n}=n\ket{n}$, where $a= (\pi_x+i\pi_y)/\sqrt{2B}$ and $[a,a^\dagger]=1$ 
at large finite $n$. 
At small $B$, the low-energy spectrum has the Dirac LL form $E_n \sim \sqrt{Bn}$. As 
$B$ is increased, pairs of LLs oscillate about each other in a `serpentine' manner, becoming degenerate at a series of crossing points absent for $v>v_c$. 
All crossings occur at a set of `magic fields', 
\begin{equation}\label{eq:Bm}
\!\!\!B_\m = \frac{m (\Delta -  m v^2) }{2 \m + 1} = 
\frac{1}{\pi\gamma^2}\frac{ \mathcal{S}_0}{(2\m+1)}, \,\, \m =0,1 ,\ldots,
\end{equation}
whence  we obtain \eqref{eq:ZLL}. The final set of crossings occurs at $B_0 = k_0^2/\gamma^{2}$, beyond which $E_n\sim Bn$ as for the unhybridized bands. The central pair of LLs that oscillate about $E=0$ are well-separated from other LLs.

In a rotated Pauli basis $\sigma^\alpha = e^{\frac{i\pi  \tau^y}{4}}\tau^\alpha e^{-\frac{i\pi \tau^y}{4}}$, we have
\begin{equation}
H  
= \left(\frac{|\bm{\pi}|^2}{2m} - \frac{\Delta}{2}\right)\sigma^x + v \pi_y \sigma^y =
\begin{pmatrix}
0 & h^\dagger\\
h & 0
\end{pmatrix},
\label{eq:H}\end{equation}
where  $h$ is the non-hermitian Hamitonian for a 1D harmonic oscillator in a constant imaginary vector potential,
\begin{equation}
h=\omega \left[a^\dagger a -\delta + \eta(a-a^\dagger)\right]
\equiv \omega(A^{\dagger}_+ A_- - \Gamma),
\end{equation}
with  $\omega= \frac{B v_c}{k_0}$, $\delta=\frac{k_0^2}{2 B}-\frac{1}{2}$, $\eta =\frac{v}{v_c}\frac{k_0}{\sqrt{2 B}}$,
$\Gamma= \frac{k_0^2}{2\gamma^2 B} -\frac{1}{2}$.
The shifted ladder operators $A_{\pm} = a\pm \eta$ are related to $a$ via a similarity transformation by $W=W^\dagger=e^{\eta (a+a^\dagger)}$,
\begin{equation}
A_{+}^\dagger = W a^\dagger W^{-1},\,A_- = W a W^{-1}.
\end{equation}
$h$ is diagonalized by right and left eigenvectors $\ket{w_n} = W\ket{n}$ and $\bra{\bar{w}_n}=\bra{n}W^{-1}$ with eigenvalues $\lambda_n = \omega(n-\Gamma)$, 
that are biorthogonal: $\braket{\bar{w}_{n}|w_{n^\prime}}=\delta_{n n^\prime}$. We observe that if $h^\dagger h\ket{\phi_i} = E_i^2\ket{\phi_i}$ with $E_i\neq 0$, then the states $\ket{\psi_{i,\pm}} = \left(\ket{\phi_i}, \pm \frac{h}{E_i}\ket{\phi_i}\right)$ are eigenstates of $H$ with eigenvalues $\pm E_i$. 
Since the parity operator $\mathcal{P}= (-1)^{a^\dagger a}\sigma^x \equiv P_a \sigma^x$ commutes with $H$,   $\mathcal{P}\ket{\psi_{i,\pm}} =  \left(\pm P_a\frac{h}{E_i}\ket{{\phi}_i}, P_a\ket{{\phi}_i}\right)$ is also an eigenvector of $H$ with eigenvalue $\pm E_i$. For generic $B\neq B_N$, $\mathcal{P}$ does not enforce any  degeneracies, and there are no $E=0$ solutions.

At magic fields $B=B_N>0$, which exist only if $v < v_c$, additional structure emerges. First, $\Gamma=N$, so that
$\ket{\psi^\Nlabel_0}=(\ket{\phi_0},0)$ and $\ket{\bar{\psi}^\Nlabel_0} = (0,\ket{\bar{\phi}_0})$,
where 
\begin{equation}
\ket{\phi_0}=\mathcal{N}_N^{-\frac{1}{2}}\ket{w_N}, \;
\ket{\bar{\phi}_0}=\mathcal{N}_N^{-\frac{1}{2}}\ket{\bar{w}_N}, 
\end{equation}
with $\mathcal{N}_\m=\braket{w_\m|w_\m}$, are exact $E=0$ eigenstates of $H$. 
Furthermore, $h^\dagger h$ only mixes $\ket{w_n}$ with $\ket{w_{n\pm 1}}$ and annihilates $\ket{w_N}$,
and hence has a decoupled subspace spanned by $\{w_{n\leq N}\}$. 
Apart from the zero-energy states, this gives rise to $2N$ energy levels $\ket{\psi_{i,\pm}}, 
i=1, 2,\ldots N$ with $|E_i| \neq 0$  expressible in terms of a {\it finite} sum of $\ket{w_n}$, 
via  $\ket{\phi_i^\Nlabel} = \sum_{n=0}^N \phi^\Nlabel_{i, n} \ket{{w}_n}$. We now observe that   $P_a h  = h^\dagger P_a$, $P_a W P_a = W^{-1}$, and hence $P_a \ket{w_n} =  (-1)^n \ket{\bar{w}_n}$; therefore  $P_a h \ket{{\phi}^\Nlabel_i} = h^\dagger P_a\ket{{\phi}^\Nlabel_i} \equiv \sum_{n=0}^N \bar{\phi}^\Nlabel_{i, n} \ket{\bar{w}_n}$ is a finite sum of $\ket{\bar{w}_n}$.  This forces $\ket{{\phi}^\Nlabel_i}$ and $P_a h \ket{{\phi}^\Nlabel_i}$ to be linearly independent~\cite{supp}:  this follows since linear dependence  requires $\langle \bar{w}_m|P_a h  \ket{{\phi}^\Nlabel_i} \propto \langle \bar{w}_m  \ket{{\phi}^\Nlabel_i}=0$ for $m>N$ (due to biorthogonality), but since  $\langle \bar{w}_m | \bar{w}_n\rangle\neq 0$ generically, this overconstrains the $\bar{\phi}^\Nlabel_{i, n}$ and forces them to vanish, leading to a contradiction. 
Since  $\ket{\phi_i^\Nlabel}$ and  $P_a h\ket{{\phi}^\Nlabel_i}$ are linearly independent, so are  $\ket{\psi_{i,\pm}}$ and $\mathcal{P}\ket{\psi_{i,\pm}}$. Thus, each $\pm E_i$ level is twofold degenerate, with orthogonal eigenstates obtained by projecting $\ket{\psi_{i\pm}}$ onto the even and odd parity sectors; including the $E=0$ states there are thus $2N+1$ pairwise LL crossings at $B_N$. As $B$ is increased above $B_N$, the LLs split linearly with  same-parity LLs from  adjacent  energies  approaching each other and undergoing avoided crossings. The $2N-1$   LL pairs closest to $E=0$   again show crossings  at $B_{N-1}$. 
The LLs thus oscillate about each other, forming serpentine  pairs whose final crossing occurs at successively lower $B$ for increasing $|E|$ (Fig.~\ref{fig:energy}). Since this  structure relies on repulsion between adjacent LL pairs, it cannot be obtained by working perturbatively within each pair.  
The crossings are all observed to occur within the band overlap window $|E|<\Delta/2$, and are a manifestation of the  crossover between the distinct LL structures  of the Dirac points and the parabolic band inversions. 
For  $v>v_c$,  $\Gamma<0$ for all $B$ and so there are no magic fields, and the serpentine structure is lost. 

We can link the unusual structure of $H$ to its membership in a class of  quasi-exactly solvable models~\cite{Turbiner1988}.
Specifically, it can be mapped to an `analytic continuation' of the Rabi model~\cite{supp}, which is known to have deep analytical structure~\cite{Braak2011,Xie_2017}.
Finally, we discuss the effect of various perturbations on ZQOs in the supplemental material~\cite{supp}.


\emph{Energy envelope.---} We estimate the ZQO envelope using  perturbation theory near $B_\m$,
yielding definite-parity eigenstates $\frac{1}{\sqrt{2}}(\ket{\psi_0}\pm \ket{\bar{\psi}_0})$
with approximate energies
\begin{equation}
E_{\pm}(B) \approx \pm \Eenv(B) \cos(\pi B_0/2 B) \label{eq:EpmB}
\end{equation}
to $O(|B-B_\m|^2)$ near any $B_\m$, where 
$\Eenv(B_\m)=B_\m v_c/(\pi k_0 \mathcal{N}_\m)$
is analytically continued to all $B$, and  $\mathcal{N}_n = e^{-2\eta^2}{}_1 F_1(1+n;1;4\eta^2)$ with $_1F_1$ the hypergeometric function.
Asymptotically, for $B\ll B_0$ we have
\begin{equation}
\Eenv(B)\approx \begin{cases}
B v_c/[\pi k_0 I_0(B^\prime/B)], & B\not\ll (\gamma^2-1) B_0\\
\sqrt{4 B v v_c/\pi}e^{-B^*/B}, & B \ll (\gamma^2-1) B_0
\end{cases}\label{eq:Eenv}
\end{equation}
where $I_0(x)$ is a Bessel function, $B^\prime=2 B_0\sqrt{\gamma^2-1}$, and 
$B^* = B_0 \{\gamma\sqrt{\gamma^2-1} - \log(\gamma - \sqrt{\gamma^2-1})\}$~\cite{supp}. 
ZQOs depend exponentially on $B$ for $B\ll B^\prime$ and saturate for  $B>B^\prime$. For $v\ll v_c$, $B'<B_0$ so there is a finite window where ZQOs are especially amenable to detection (though they are likely observable even in the exponential regime, which describes the entirety of Fig.~\ref{fig:energy}).
As $v$ increases, this window shrinks, vanishing at $v= v_c/\sqrt{5}$ when $B'=B_0$. 
For $v_c/\sqrt{5}\leq v<v_c$, ZQOs are always in the exponential regime, and are absent for $v\geq v_c$. The exponential onset  in $B$ is a signature of ZQOs. 

\emph{Non-LK Temperature Dependence.---} Since ZQOs arise from the motion of just two states, we expect distinctive  thermodynamic signatures at charge neutrality and low but finite temperature $T=\beta^{-1}$, controlled by the low-energy density of states per unit area~\cite{Zhang2016}
\begin{equation}
\!\!\! D_\beta(B) = -\int d E n_F^\prime(E) \rho(E) \approx \frac{\beta B/(4\pi)}{\cosh (\beta E_{\pm}(B)/2)^2},
\end{equation}
where  $n_F^\prime(E)=-\beta e^{\beta E}/(1+e^{\beta E})^{2}$ is the derivative of the Fermi function, $\rho$ is the single particle density of states, and we have taken into account the LL degeneracy per unit area $B/2\pi$.
The final form is valid at temperatures where only the two central states $E_{\pm}(B)$ contribute, and leads to the non-LK behaviour \eqref{eq:RZLL}  in the oscillation magnitude.
In the metallic limit $v=0$, or at very high temperatures, the central two states can no longer be treated as separate from the remaining spectrum and the conventional LK form~\eqref{eq:LK} is restored.

\emph{Topological crystalline insulators. ---} Having explained the origin of ZQOs in a solvable model and argued that they survive on relaxing  its simplifying assumptions (and in the tight-binding limit~\cite{supp}),  we now turn to identifying experimental settings where they may be observable. 
Although the models discussed thus far are quite fine-tuned for quasi-exact solvability, the physics of ZQOs should be relatively universal in Dirac systems with weakly-gapped nodal rings.
As a concrete example, consider the 2D surface of a 3D topological crystalline insulator~\cite{Fu2011} (TCI) with mirror symmetry (a description that encompasses the SnTe material class), described by the $k\cdot p$ Hamiltonian~\cite{Liu2013}
\begin{equation}
H_{{T}} = v_T( k_x s^y - k_y s^x) + m_T \tau^x + \delta_T s^x \tau^y  .
\end{equation}
Although very different from $H$, the low-energy dispersion of $H_{{T}}$ also hosts two Dirac cones connected via a weakly-gapped nodal ring
~\footnote{Ref~\cite{Serbyn2014} observed (but did not discuss) ZQOs in this system.}.
The positions of the two nodal points and their velocities are
$k_0=\sqrt{\delta_T^2+m_T^2}/v_T,\; v_x=v_T,\; v_y=\delta_T/k_0$. 
The zero-energy eigenstates of $H_T$ can be determined exactly and appear at fields $B_{T,N}=m_T^2/(2 v_T^2 \m)$ for positive integers $\m$.  
The oscillation magnitude $\mathcal{E}_T(B)$ displays the same qualitative behavior as $\mathcal{E}(B)$~\cite{supp}. 
The parameters of $H_{T}$ have been  experimentally determined in a number of materials~\cite{Liu2013,Hsieh2012,Okada2013,Wang2013,Assaf2016}.
ZQOs occur for $B < B_{T,1}$, and depend exponentially on inverse field when $B\ll B^\prime_T=2\delta_T m_T/v_T^2$, from \eqref{eq:Eenv}.
As a typical example,  parameters relevant to SnTe ~\cite{Liu2013,Hsieh2012} give
$(B^\prime_T,B_{T,1})\approx(\SI{54}{\tesla},\SI{52}{\tesla})$ with the envelope function $\Eenv_T(\SI{15}{\tesla})\approx\SI{1}~\text{K}$~\cite{supp}. 
This places the corresponding ZQOs in the exponential regime, and are observable within the ranges of fields and temperatures currently achievable in experiments using pulsed magnetic fields~\cite{QOCuprates}.
The parameters $m_T$ and $\delta_T$ are tunable by strain~\cite{Serbyn2014}: even a small change can have a significant effect on the ZQOs due to the exponential dependence.
ZQOs can therefore serve as a sensitive probe of TCI surface states.

\begin{figure}
\includegraphics[width=0.5\textwidth]{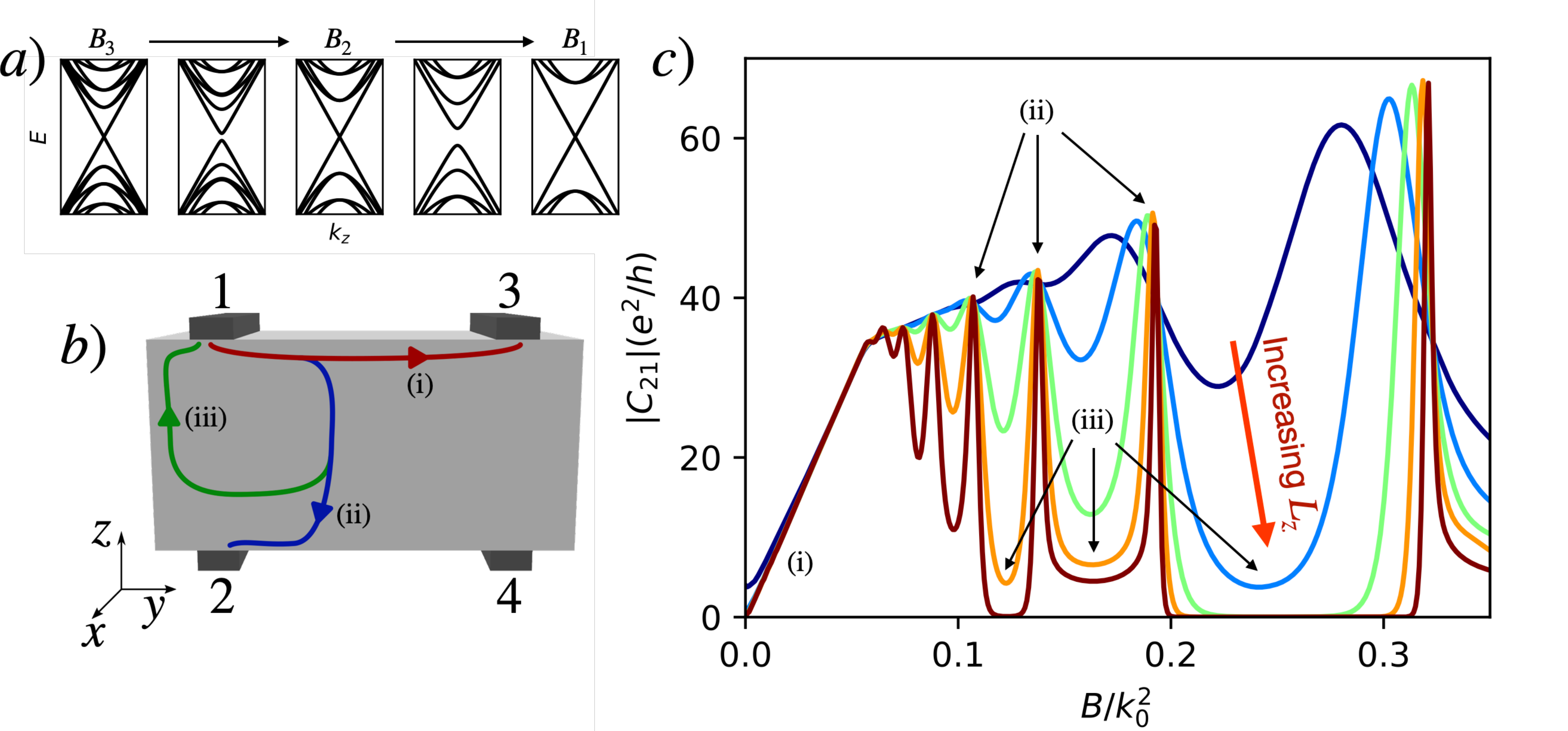}
\caption{
$(a)$ The LL dispersion of $H_\mathrm{Weyl}$ at, and in between, magic fields $B_3,B_2$, and $B_1$ for $v=v_c/6$, illustrating the $1/B$-periodic gap closings of the chiral LLs resulting from ZQOs.
$(b)$ A proposed experimental set-up for observing ZQOs in 3D.  Four leads are attached to a bulk Weyl semimetal.  $(c)$ Numerical results for the $C_{21}$ element of the conductance matrix shows clear peaks at the magic fields.
}
\label{fig:weyl}
\end{figure}

\emph{Three dimensions. ---}
The physics of ZQOs also generalize to 3D.
A prototypical model for a Weyl semimetal with broken $\mathcal{T}$ symmetry~\cite{Armitage2018}, 
\begin{equation}
H_{\mathrm{Weyl}}(\bm{k}) = \left(\frac{|\bm{k}|^2}{2m}-\frac{\Delta}{2}\right)\sigma^x + vk_y \sigma^y + w k_z \sigma^z,
\label{eq:Hweyl}
\end{equation}
coincides with \eqref{eq:H0} (up to a mass term) at each $k_z$, and therefore exhibits ZQOs under a magnetic field $B\hat{z}$ for $v<v_c$.
The condition $v < v_c$ is naturally satisfied in Weyl semimetals which are proximate to a nodal line semimetal limit $v=0$ where the two nodes form a nodal ring, for example in scenarios where the protecting symmetries of the Weyl ring are weakly broken~\cite{Chan2016,Fang2016}.
The presence of zero-modes at $k_z=0$ in a magnetic field corresponds to the existence of gapless bulk chiral LLs which propagate along $\pm\hat{z}$.
Hence, ZQOs correspond to a periodic opening and vanishing of the gap between these chiral LLs, as illustrated in Fig~\ref{fig:weyl}a.  
Since gaplessness of these LLs is crucial to magnetotransport effects linked to the chiral anomaly and Fermi arcs~\cite{Potter2014,PhysRevB.88.104412,PhysRevX.4.031035, PhysRevX.5.041046, HOSUR2013857}, a striking implication of ZQOs is that such phenomena will also experience $1/B$-periodic revivals.
As a demonstration, Fig~\ref{fig:weyl}b shows a transport measurement capable of probing ZQOs: 
four leads labeled by $i=1\dots 4$ are connected to a bulk sample and the conductance matrix $C_{ij}$, defined in terms of the lead current and voltage via $I_i = C_{ij}V_j$, is measured.
Fig~\ref{fig:weyl}c shows that $C_{21}$, computed numerically at $T=0$ for a discretized model, is highly sensitive to ZQOs and can be understood as follows.
At zero and low fields, electrons from lead $1$ flow mainly into Fermi arc surface states, which drives current $1\rightarrow 3$ as indicated by (i).
At higher fields, electrons are transferred to the bulk chiral modes before making it to lead $3$~\cite{Potter2014}. 
When these chiral modes are perfect (i.e. gapless at $B=B_N$), electrons fully traverse the bulk and drives current $1\rightarrow 2$ (ii).
Otherwise, electrons are only able to traverse a finite distance into the bulk before turning back resulting in no current (iii).
Indeed, $C_{21}$ demonstrates clear peaks at each magic field which become sharper with increasing thickness $L_z$.
Details and discussion of the numerical calculation, performed using the Kwant~\cite{kwant} code, is available in the supplemental material~\cite{supp}.

\emph{Conclusions.---}
While we postpone a detailed analysis~\cite{TDunpub}, we observe in closing that in the $B\to 0$ limit, ZQOs may be understood by considering  $E=0$ semiclassical tunnelling trajectories in the BZ between the Dirac points at $\pm \bm{k}_0$. These capture instanton events that generate repulsion between 
the zero-energy LLs of the Dirac points;
they have a complex action, leading to a tunneling matrix element whose phase (amplitude) depends on $\pi B_0$ ($B^*$).
Summing over tunneling events~\cite{rajaraman1982solitons}  and recalling that $B$ plays the role of $\hbar$ in the semiclassical expansion yields a result consistent with the  $B\to 0$ limit of \eqref{eq:EpmB}.  Intuitively, ZQOs emerge when optimal tunneling trajectories linking distinct $B=0$ Dirac points in the BZ  
involve an excursion to $k_y\neq 0$ acquiring
an Aharanov-Bohm phase and its associated interference effects, which  vanish for $v\geq v_c$ when these trajectories are fixed at $k_y=0$. (Similar considerations yield  QOs in narrow-gap insulators, but since these lack  zero-energy LLs,  it is more natural to view semiclassical effects as modulating  the LL gap at $E_F$ rather than generating LL repulsion.) 
This places ZQOs on conceptually similar footing with `tunneling interference' effects~\cite{JK1,
JK2,
Sharpee,
SpinTun1,
SpinTun2,
SpinAB} and  indicates that these ideas are broadly applicable. 
Finally, as a quasi-exactly-solvable model amenable to semiclassics, \eqref{eq:H0}  is potentially  noteworthy in the context of `resurgent asymptotics' in quantum mechanics~\cite{JentschuraZinnJustin,DunneUnsalWKB,dorigoni2015introduction,DunneUnsal}.

\begin{acknowledgments}
We thank B. Assaf, B.A. Bernevig, A.I. Coldea, F.H.L. Essler,  and L. Fu for discussions. We are especially grateful to N. Cooper for clarifying aspects of~\cite{Knolle2015,Knolle2017a} and for a stimulating conversation on semiclassical trajectories in quantum oscillations.  We acknowledge support from the European Research Council under the European Union Horizon 2020 Research and Innovation Programme, Grant Agreement No. 804213-TMCS (YHK, SAP). Additional support was provided by the Gordon and Betty Moore Foundation through Grant GBMF8685 towards the Princeton theory program.
\end{acknowledgments}
\bibliography{references}{}

\begin{thebibliography}{62}%
\makeatletter
\providecommand \@ifxundefined [1]{%
 \@ifx{#1\undefined}
}%
\providecommand \@ifnum [1]{%
 \ifnum #1\expandafter \@firstoftwo
 \else \expandafter \@secondoftwo
 \fi
}%
\providecommand \@ifx [1]{%
 \ifx #1\expandafter \@firstoftwo
 \else \expandafter \@secondoftwo
 \fi
}%
\providecommand \natexlab [1]{#1}%
\providecommand \enquote  [1]{``#1''}%
\providecommand \bibnamefont  [1]{#1}%
\providecommand \bibfnamefont [1]{#1}%
\providecommand \citenamefont [1]{#1}%
\providecommand \href@noop [0]{\@secondoftwo}%
\providecommand \href [0]{\begingroup \@sanitize@url \@href}%
\providecommand \@href[1]{\@@startlink{#1}\@@href}%
\providecommand \@@href[1]{\endgroup#1\@@endlink}%
\providecommand \@sanitize@url [0]{\catcode `\\12\catcode `\$12\catcode
  `\&12\catcode `\#12\catcode `\^12\catcode `\_12\catcode `\%12\relax}%
\providecommand \@@startlink[1]{}%
\providecommand \@@endlink[0]{}%
\providecommand \url  [0]{\begingroup\@sanitize@url \@url }%
\providecommand \@url [1]{\endgroup\@href {#1}{\urlprefix }}%
\providecommand \urlprefix  [0]{URL }%
\providecommand \Eprint [0]{\href }%
\providecommand \doibase [0]{http://dx.doi.org/}%
\providecommand \selectlanguage [0]{\@gobble}%
\providecommand \bibinfo  [0]{\@secondoftwo}%
\providecommand \bibfield  [0]{\@secondoftwo}%
\providecommand \translation [1]{[#1]}%
\providecommand \BibitemOpen [0]{}%
\providecommand \bibitemStop [0]{}%
\providecommand \bibitemNoStop [0]{.\EOS\space}%
\providecommand \EOS [0]{\spacefactor3000\relax}%
\providecommand \BibitemShut  [1]{\csname bibitem#1\endcsname}%
\let\auto@bib@innerbib\@empty
\bibitem [{\citenamefont {Shubnikov}\ and\ \citenamefont {{W.J. de
  Haas}}(1930)}]{SdH}%
  \BibitemOpen
  \bibfield  {author} {\bibinfo {author} {\bibfnamefont {L.}~\bibnamefont
  {Shubnikov}}\ and\ \bibinfo {author} {\bibnamefont {{W.J. de Haas}}},\
  }\href@noop {} {\bibfield  {journal} {\bibinfo  {journal} {Proc. Neth. R.
  Acad. Sci.}\ }\textbf {\bibinfo {volume} {33}},\ \bibinfo {pages} {130}
  (\bibinfo {year} {1930})}\BibitemShut {NoStop}%
\bibitem [{\citenamefont {{W.J. de Haas}}\ and\ \citenamefont {{P. M. van
  Alphen}}(1930)}]{dHvA}%
  \BibitemOpen
  \bibfield  {author} {\bibinfo {author} {\bibnamefont {{W.J. de Haas}}}\ and\
  \bibinfo {author} {\bibnamefont {{P. M. van Alphen}}},\ }\href@noop {}
  {\bibfield  {journal} {\bibinfo  {journal} {Proc. Neth. R. Acad. Sci.}\
  }\textbf {\bibinfo {volume} {33}},\ \bibinfo {pages} {1106} (\bibinfo {year}
  {1930})}\BibitemShut {NoStop}%
\bibitem [{\citenamefont {Shoenberg}(1984)}]{shoenberg_1984}%
  \BibitemOpen
  \bibfield  {author} {\bibinfo {author} {\bibfnamefont {D.}~\bibnamefont
  {Shoenberg}},\ }\href {\doibase 10.1017/CBO9780511897870} {\emph {\bibinfo
  {title} {Magnetic Oscillations in Metals}}},\ Cambridge Monographs on
  Physics\ (\bibinfo  {publisher} {Cambridge University Press},\ \bibinfo
  {year} {1984})\BibitemShut {NoStop}%
\bibitem [{\citenamefont {Onsager}(1952)}]{Onsager}%
  \BibitemOpen
  \bibfield  {author} {\bibinfo {author} {\bibfnamefont {L.}~\bibnamefont
  {Onsager}},\ }\href {\doibase 10.1080/14786440908521019} {\bibfield
  {journal} {\bibinfo  {journal} {Phil. Mag.}\ }\textbf {\bibinfo {volume}
  {43}},\ \bibinfo {pages} {1006} (\bibinfo {year} {1952})}\BibitemShut
  {NoStop}%
\bibitem [{\citenamefont {Lifshitz}\ and\ \citenamefont {Kosevich}(1956)}]{LK}%
  \BibitemOpen
  \bibfield  {author} {\bibinfo {author} {\bibfnamefont {L.}~\bibnamefont
  {Lifshitz}}\ and\ \bibinfo {author} {\bibfnamefont {A.}~\bibnamefont
  {Kosevich}},\ }\href@noop {} {\bibfield  {journal} {\bibinfo  {journal}
  {Soviet Phys. JETP}\ }\textbf {\bibinfo {volume} {2}},\ \bibinfo {pages}
  {636} (\bibinfo {year} {1956})}\BibitemShut {NoStop}%
\bibitem [{\citenamefont {Armitage}\ \emph {et~al.}(2018)\citenamefont
  {Armitage}, \citenamefont {Mele},\ and\ \citenamefont
  {Vishwanath}}]{Armitage2018}%
  \BibitemOpen
  \bibfield  {author} {\bibinfo {author} {\bibfnamefont {N.~P.}\ \bibnamefont
  {Armitage}}, \bibinfo {author} {\bibfnamefont {E.~J.}\ \bibnamefont {Mele}},
  \ and\ \bibinfo {author} {\bibfnamefont {A.}~\bibnamefont {Vishwanath}},\
  }\href {\doibase 10.1103/RevModPhys.90.015001} {\bibfield  {journal}
  {\bibinfo  {journal} {Rev. Mod. Phys.}\ }\textbf {\bibinfo {volume} {90}},\
  \bibinfo {pages} {015001} (\bibinfo {year} {2018})}\BibitemShut {NoStop}%
\bibitem [{\citenamefont {Mikitik}\ and\ \citenamefont
  {Sharlai}(1999)}]{Mikitik99}%
  \BibitemOpen
  \bibfield  {author} {\bibinfo {author} {\bibfnamefont {G.~P.}\ \bibnamefont
  {Mikitik}}\ and\ \bibinfo {author} {\bibfnamefont {Y.~V.}\ \bibnamefont
  {Sharlai}},\ }\href {\doibase 10.1103/PhysRevLett.82.2147} {\bibfield
  {journal} {\bibinfo  {journal} {Phys. Rev. Lett.}\ }\textbf {\bibinfo
  {volume} {82}},\ \bibinfo {pages} {2147} (\bibinfo {year}
  {1999})}\BibitemShut {NoStop}%
\bibitem [{\citenamefont {Alexandradinata}\ and\ \citenamefont
  {Glazman}(2017)}]{AG1}%
  \BibitemOpen
  \bibfield  {author} {\bibinfo {author} {\bibfnamefont {A.}~\bibnamefont
  {Alexandradinata}}\ and\ \bibinfo {author} {\bibfnamefont {L.}~\bibnamefont
  {Glazman}},\ }\href {\doibase 10.1103/PhysRevLett.119.256601} {\bibfield
  {journal} {\bibinfo  {journal} {Phys. Rev. Lett.}\ }\textbf {\bibinfo
  {volume} {119}},\ \bibinfo {pages} {256601} (\bibinfo {year}
  {2017})}\BibitemShut {NoStop}%
\bibitem [{\citenamefont {Alexandradinata}\ and\ \citenamefont
  {Glazman}(2018)}]{AG2}%
  \BibitemOpen
  \bibfield  {author} {\bibinfo {author} {\bibfnamefont {A.}~\bibnamefont
  {Alexandradinata}}\ and\ \bibinfo {author} {\bibfnamefont {L.}~\bibnamefont
  {Glazman}},\ }\href {\doibase 10.1103/PhysRevB.97.144422} {\bibfield
  {journal} {\bibinfo  {journal} {Phys. Rev. B}\ }\textbf {\bibinfo {volume}
  {97}},\ \bibinfo {pages} {144422} (\bibinfo {year} {2018})}\BibitemShut
  {NoStop}%
\bibitem [{\citenamefont {Balatskii}\ \emph {et~al.}(1986)\citenamefont
  {Balatskii}, \citenamefont {Volovik},\ and\ \citenamefont
  {Konyshev}}]{balatskii1986chiral}%
  \BibitemOpen
  \bibfield  {author} {\bibinfo {author} {\bibfnamefont {A.}~\bibnamefont
  {Balatskii}}, \bibinfo {author} {\bibfnamefont {G.}~\bibnamefont {Volovik}},
  \ and\ \bibinfo {author} {\bibfnamefont {A.}~\bibnamefont {Konyshev}},\
  }\href@noop {} {\bibfield  {journal} {\bibinfo  {journal} {Zh. Eksp. Teor.
  Fiz}\ }\textbf {\bibinfo {volume} {90}},\ \bibinfo {pages} {2038} (\bibinfo
  {year} {1986})}\BibitemShut {NoStop}%
\bibitem [{\citenamefont {Zhang}\ \emph {et~al.}(2017)\citenamefont {Zhang},
  \citenamefont {Xu}, \citenamefont {Wang}, \citenamefont {Lin}, \citenamefont
  {Du}, \citenamefont {Guo}, \citenamefont {Lee}, \citenamefont {Lu},
  \citenamefont {Feng}, \citenamefont {Huang}, \citenamefont {Chang},
  \citenamefont {Hsu}, \citenamefont {Liu}, \citenamefont {Lin}, \citenamefont
  {Li}, \citenamefont {Zhang}, \citenamefont {Zhang}, \citenamefont {Xie},
  \citenamefont {Neupert}, \citenamefont {Hasan}, \citenamefont {Lu},
  \citenamefont {Wang},\ and\ \citenamefont {Jia}}]{WeylAnnihilation}%
  \BibitemOpen
  \bibfield  {author} {\bibinfo {author} {\bibfnamefont {C.-L.}\ \bibnamefont
  {Zhang}}, \bibinfo {author} {\bibfnamefont {S.-Y.}\ \bibnamefont {Xu}},
  \bibinfo {author} {\bibfnamefont {C.~M.}\ \bibnamefont {Wang}}, \bibinfo
  {author} {\bibfnamefont {Z.}~\bibnamefont {Lin}}, \bibinfo {author}
  {\bibfnamefont {Z.~Z.}\ \bibnamefont {Du}}, \bibinfo {author} {\bibfnamefont
  {C.}~\bibnamefont {Guo}}, \bibinfo {author} {\bibfnamefont {C.-C.}\
  \bibnamefont {Lee}}, \bibinfo {author} {\bibfnamefont {H.}~\bibnamefont
  {Lu}}, \bibinfo {author} {\bibfnamefont {Y.}~\bibnamefont {Feng}}, \bibinfo
  {author} {\bibfnamefont {S.-M.}\ \bibnamefont {Huang}}, \bibinfo {author}
  {\bibfnamefont {G.}~\bibnamefont {Chang}}, \bibinfo {author} {\bibfnamefont
  {C.-H.}\ \bibnamefont {Hsu}}, \bibinfo {author} {\bibfnamefont
  {H.}~\bibnamefont {Liu}}, \bibinfo {author} {\bibfnamefont {H.}~\bibnamefont
  {Lin}}, \bibinfo {author} {\bibfnamefont {L.}~\bibnamefont {Li}}, \bibinfo
  {author} {\bibfnamefont {C.}~\bibnamefont {Zhang}}, \bibinfo {author}
  {\bibfnamefont {J.}~\bibnamefont {Zhang}}, \bibinfo {author} {\bibfnamefont
  {X.-C.}\ \bibnamefont {Xie}}, \bibinfo {author} {\bibfnamefont
  {T.}~\bibnamefont {Neupert}}, \bibinfo {author} {\bibfnamefont {M.~Z.}\
  \bibnamefont {Hasan}}, \bibinfo {author} {\bibfnamefont {H.-Z.}\ \bibnamefont
  {Lu}}, \bibinfo {author} {\bibfnamefont {J.}~\bibnamefont {Wang}}, \ and\
  \bibinfo {author} {\bibfnamefont {S.}~\bibnamefont {Jia}},\ }\href {\doibase
  10.1038/nphys4183} {\bibfield  {journal} {\bibinfo  {journal} {Nature
  Physics}\ }\textbf {\bibinfo {volume} {13}},\ \bibinfo {pages} {979}
  (\bibinfo {year} {2017})}\BibitemShut {NoStop}%
\bibitem [{\citenamefont {Chan}\ and\ \citenamefont {Lee}(2017)}]{ChanLee}%
  \BibitemOpen
  \bibfield  {author} {\bibinfo {author} {\bibfnamefont {C.-K.}\ \bibnamefont
  {Chan}}\ and\ \bibinfo {author} {\bibfnamefont {P.~A.}\ \bibnamefont {Lee}},\
  }\href {\doibase 10.1103/PhysRevB.96.195143} {\bibfield  {journal} {\bibinfo
  {journal} {Phys. Rev. B}\ }\textbf {\bibinfo {volume} {96}},\ \bibinfo
  {pages} {195143} (\bibinfo {year} {2017})}\BibitemShut {NoStop}%
\bibitem [{\citenamefont {Bednik}\ \emph {et~al.}(2020)\citenamefont {Bednik},
  \citenamefont {Tikhonov},\ and\ \citenamefont {Syzranov}}]{Bednik}%
  \BibitemOpen
  \bibfield  {author} {\bibinfo {author} {\bibfnamefont {G.}~\bibnamefont
  {Bednik}}, \bibinfo {author} {\bibfnamefont {K.~S.}\ \bibnamefont
  {Tikhonov}}, \ and\ \bibinfo {author} {\bibfnamefont {S.~V.}\ \bibnamefont
  {Syzranov}},\ }\href {\doibase 10.1103/PhysRevResearch.2.023124} {\bibfield
  {journal} {\bibinfo  {journal} {Phys. Rev. Research}\ }\textbf {\bibinfo
  {volume} {2}},\ \bibinfo {pages} {023124} (\bibinfo {year}
  {2020})}\BibitemShut {NoStop}%
\bibitem [{\citenamefont {Wang}\ \emph {et~al.}(2020)\citenamefont {Wang},
  \citenamefont {Lu},\ and\ \citenamefont {Xie}}]{Wang2020}%
  \BibitemOpen
  \bibfield  {author} {\bibinfo {author} {\bibfnamefont {C.~M.}\ \bibnamefont
  {Wang}}, \bibinfo {author} {\bibfnamefont {H.-Z.}\ \bibnamefont {Lu}}, \ and\
  \bibinfo {author} {\bibfnamefont {X.~C.}\ \bibnamefont {Xie}},\ }\href
  {\doibase 10.1103/PhysRevB.102.041204} {\bibfield  {journal} {\bibinfo
  {journal} {Phys. Rev. B}\ }\textbf {\bibinfo {volume} {102}},\ \bibinfo
  {pages} {041204} (\bibinfo {year} {2020})}\BibitemShut {NoStop}%
\bibitem [{\citenamefont {Potter}\ \emph {et~al.}(2014)\citenamefont {Potter},
  \citenamefont {Kimchi},\ and\ \citenamefont {Vishwanath}}]{Potter2014}%
  \BibitemOpen
  \bibfield  {author} {\bibinfo {author} {\bibfnamefont {A.~C.}\ \bibnamefont
  {Potter}}, \bibinfo {author} {\bibfnamefont {I.}~\bibnamefont {Kimchi}}, \
  and\ \bibinfo {author} {\bibfnamefont {A.}~\bibnamefont {Vishwanath}},\
  }\href {\doibase 10.1038/ncomms6161} {\bibfield  {journal} {\bibinfo
  {journal} {Nature Communications}\ }\textbf {\bibinfo {volume} {5}},\
  \bibinfo {pages} {5161} (\bibinfo {year} {2014})}\BibitemShut {NoStop}%
\bibitem [{\citenamefont {Knolle}\ and\ \citenamefont
  {Cooper}(2015)}]{Knolle2015}%
  \BibitemOpen
  \bibfield  {author} {\bibinfo {author} {\bibfnamefont {J.}~\bibnamefont
  {Knolle}}\ and\ \bibinfo {author} {\bibfnamefont {N.~R.}\ \bibnamefont
  {Cooper}},\ }\href {\doibase 10.1103/PhysRevLett.115.146401} {\bibfield
  {journal} {\bibinfo  {journal} {Phys. Rev. Lett.}\ }\textbf {\bibinfo
  {volume} {115}},\ \bibinfo {pages} {146401} (\bibinfo {year}
  {2015})}\BibitemShut {NoStop}%
\bibitem [{\citenamefont {Knolle}\ and\ \citenamefont
  {Cooper}(2017)}]{Knolle2017a}%
  \BibitemOpen
  \bibfield  {author} {\bibinfo {author} {\bibfnamefont {J.}~\bibnamefont
  {Knolle}}\ and\ \bibinfo {author} {\bibfnamefont {N.~R.}\ \bibnamefont
  {Cooper}},\ }\href {\doibase 10.1103/PhysRevLett.118.176801} {\bibfield
  {journal} {\bibinfo  {journal} {Phys. Rev. Lett.}\ }\textbf {\bibinfo
  {volume} {118}},\ \bibinfo {pages} {176801} (\bibinfo {year}
  {2017})}\BibitemShut {NoStop}%
\bibitem [{\citenamefont {Zhang}\ \emph {et~al.}(2016)\citenamefont {Zhang},
  \citenamefont {Song},\ and\ \citenamefont {Wang}}]{Zhang2016}%
  \BibitemOpen
  \bibfield  {author} {\bibinfo {author} {\bibfnamefont {L.}~\bibnamefont
  {Zhang}}, \bibinfo {author} {\bibfnamefont {X.-Y.}\ \bibnamefont {Song}}, \
  and\ \bibinfo {author} {\bibfnamefont {F.}~\bibnamefont {Wang}},\ }\href
  {\doibase 10.1103/PhysRevLett.116.046404} {\bibfield  {journal} {\bibinfo
  {journal} {Phys. Rev. Lett.}\ }\textbf {\bibinfo {volume} {116}},\ \bibinfo
  {pages} {046404} (\bibinfo {year} {2016})}\BibitemShut {NoStop}%
\bibitem [{\citenamefont {Pal}\ \emph {et~al.}(2016)\citenamefont {Pal},
  \citenamefont {Pi{\'{e}}chon}, \citenamefont {Fuchs}, \citenamefont
  {Goerbig},\ and\ \citenamefont {Montambaux}}]{Pal2016}%
  \BibitemOpen
  \bibfield  {author} {\bibinfo {author} {\bibfnamefont {H.~K.}\ \bibnamefont
  {Pal}}, \bibinfo {author} {\bibfnamefont {F.}~\bibnamefont {Pi{\'{e}}chon}},
  \bibinfo {author} {\bibfnamefont {J.-N.}\ \bibnamefont {Fuchs}}, \bibinfo
  {author} {\bibfnamefont {M.}~\bibnamefont {Goerbig}}, \ and\ \bibinfo
  {author} {\bibfnamefont {G.}~\bibnamefont {Montambaux}},\ }\href {\doibase
  10.1103/PhysRevB.94.125140} {\bibfield  {journal} {\bibinfo  {journal} {Phys.
  Rev. B}\ }\textbf {\bibinfo {volume} {94}},\ \bibinfo {pages} {125140}
  (\bibinfo {year} {2016})}\BibitemShut {NoStop}%
\bibitem [{\citenamefont {Pal}(2017)}]{Pal2017}%
  \BibitemOpen
  \bibfield  {author} {\bibinfo {author} {\bibfnamefont {H.~K.}\ \bibnamefont
  {Pal}},\ }\href {\doibase 10.1103/PhysRevB.96.235121} {\bibfield  {journal}
  {\bibinfo  {journal} {Phys. Rev. B}\ }\textbf {\bibinfo {volume} {96}},\
  \bibinfo {pages} {235121} (\bibinfo {year} {2017})}\BibitemShut {NoStop}%
\bibitem [{\citenamefont {Ram}\ and\ \citenamefont {Kumar}(2017)}]{Ram2017}%
  \BibitemOpen
  \bibfield  {author} {\bibinfo {author} {\bibfnamefont {P.}~\bibnamefont
  {Ram}}\ and\ \bibinfo {author} {\bibfnamefont {B.}~\bibnamefont {Kumar}},\
  }\href {\doibase 10.1103/PhysRevB.96.075115} {\bibfield  {journal} {\bibinfo
  {journal} {Phys. Rev. B}\ }\textbf {\bibinfo {volume} {96}},\ \bibinfo
  {pages} {075115} (\bibinfo {year} {2017})}\BibitemShut {NoStop}%
\bibitem [{\citenamefont {Ram}\ and\ \citenamefont {Kumar}(2019)}]{Ram2019}%
  \BibitemOpen
  \bibfield  {author} {\bibinfo {author} {\bibfnamefont {P.}~\bibnamefont
  {Ram}}\ and\ \bibinfo {author} {\bibfnamefont {B.}~\bibnamefont {Kumar}},\
  }\href {\doibase 10.1103/PhysRevB.99.235130} {\bibfield  {journal} {\bibinfo
  {journal} {Phys. Rev. B}\ }\textbf {\bibinfo {volume} {99}},\ \bibinfo
  {pages} {235130} (\bibinfo {year} {2019})}\BibitemShut {NoStop}%
\bibitem [{\citenamefont {Grubinskas}\ and\ \citenamefont
  {Fritz}(2018)}]{Grubinskas2018}%
  \BibitemOpen
  \bibfield  {author} {\bibinfo {author} {\bibfnamefont {S.}~\bibnamefont
  {Grubinskas}}\ and\ \bibinfo {author} {\bibfnamefont {L.}~\bibnamefont
  {Fritz}},\ }\href {\doibase 10.1103/PhysRevB.97.115202} {\bibfield  {journal}
  {\bibinfo  {journal} {Phys. Rev. B}\ }\textbf {\bibinfo {volume} {97}},\
  \bibinfo {pages} {115202} (\bibinfo {year} {2018})}\BibitemShut {NoStop}%
\bibitem [{\citenamefont {Shen}\ and\ \citenamefont {Fu}(2018)}]{Shen2018}%
  \BibitemOpen
  \bibfield  {author} {\bibinfo {author} {\bibfnamefont {H.}~\bibnamefont
  {Shen}}\ and\ \bibinfo {author} {\bibfnamefont {L.}~\bibnamefont {Fu}},\
  }\href {\doibase 10.1103/PhysRevLett.121.026403} {\bibfield  {journal}
  {\bibinfo  {journal} {Phys. Rev. Lett.}\ }\textbf {\bibinfo {volume} {121}},\
  \bibinfo {pages} {026403} (\bibinfo {year} {2018})}\BibitemShut {NoStop}%
\bibitem [{\citenamefont {Lu}\ \emph {et~al.}(2020)\citenamefont {Lu},
  \citenamefont {Chou}, \citenamefont {Chung}, \citenamefont {Lee},\ and\
  \citenamefont {Mou}}]{Lu2020}%
  \BibitemOpen
  \bibfield  {author} {\bibinfo {author} {\bibfnamefont {Y.-W.}\ \bibnamefont
  {Lu}}, \bibinfo {author} {\bibfnamefont {P.-H.}\ \bibnamefont {Chou}},
  \bibinfo {author} {\bibfnamefont {C.-H.}\ \bibnamefont {Chung}}, \bibinfo
  {author} {\bibfnamefont {T.-K.}\ \bibnamefont {Lee}}, \ and\ \bibinfo
  {author} {\bibfnamefont {C.-Y.}\ \bibnamefont {Mou}},\ }\href {\doibase
  10.1103/PhysRevB.101.115102} {\bibfield  {journal} {\bibinfo  {journal}
  {Phys. Rev. B}\ }\textbf {\bibinfo {volume} {101}},\ \bibinfo {pages}
  {115102} (\bibinfo {year} {2020})}\BibitemShut {NoStop}%
\bibitem [{\citenamefont {Falkovsky}(2011)}]{Falkovsky2011}%
  \BibitemOpen
  \bibfield  {author} {\bibinfo {author} {\bibfnamefont {L.~A.}\ \bibnamefont
  {Falkovsky}},\ }\href {\doibase 10.1063/1.3670023} {\bibfield  {journal}
  {\bibinfo  {journal} {Low Temperature Physics}\ }\textbf {\bibinfo {volume}
  {37}},\ \bibinfo {pages} {815} (\bibinfo {year} {2011})}\BibitemShut
  {NoStop}%
\bibitem [{\citenamefont {Alisultanov}(2016)}]{Alisultanov2016}%
  \BibitemOpen
  \bibfield  {author} {\bibinfo {author} {\bibfnamefont {Z.~Z.}\ \bibnamefont
  {Alisultanov}},\ }\href {\doibase 10.1134/S0021364016150042} {\bibfield
  {journal} {\bibinfo  {journal} {JETP Letters}\ }\textbf {\bibinfo {volume}
  {104}},\ \bibinfo {pages} {188} (\bibinfo {year} {2016})}\BibitemShut
  {NoStop}%
\bibitem [{sup()}]{supp}%
  \BibitemOpen
  \href@noop {} {}\bibinfo {note} {See Supplemental Material for additional
  details of the theoretical analysis, including (i) various analytic
  properties of the exact solution; (ii) TCI surface state exact solution;
  (iii) mapping to the Rabi model; (iv) a tight-binding model realization; (v)
  details regarding Weyl semimetal conductivity calculations; and (vi) details
  of the non-LK temperature dependence; which includes
  Refs~\onlinecite{HaldaneModel_fromsupp,Kane2005_fromsupp,DLMF_fromsupp}}\BibitemShut
  {NoStop}%
\bibitem [{\citenamefont {Turbiner}(1988)}]{Turbiner1988}%
  \BibitemOpen
  \bibfield  {author} {\bibinfo {author} {\bibfnamefont {A.~V.}\ \bibnamefont
  {Turbiner}},\ }\href {\doibase 10.1007/BF01466727} {\bibfield  {journal}
  {\bibinfo  {journal} {Communications in Mathematical Physics}\ }\textbf
  {\bibinfo {volume} {118}},\ \bibinfo {pages} {467} (\bibinfo {year}
  {1988})}\BibitemShut {NoStop}%
\bibitem [{\citenamefont {Braak}(2011)}]{Braak2011}%
  \BibitemOpen
  \bibfield  {author} {\bibinfo {author} {\bibfnamefont {D.}~\bibnamefont
  {Braak}},\ }\href {\doibase 10.1103/PhysRevLett.107.100401} {\bibfield
  {journal} {\bibinfo  {journal} {Phys. Rev. Lett.}\ }\textbf {\bibinfo
  {volume} {107}},\ \bibinfo {pages} {100401} (\bibinfo {year}
  {2011})}\BibitemShut {NoStop}%
\bibitem [{\citenamefont {Xie}\ \emph {et~al.}(2017)\citenamefont {Xie},
  \citenamefont {Zhong}, \citenamefont {Batchelor},\ and\ \citenamefont
  {Lee}}]{Xie_2017}%
  \BibitemOpen
  \bibfield  {author} {\bibinfo {author} {\bibfnamefont {Q.}~\bibnamefont
  {Xie}}, \bibinfo {author} {\bibfnamefont {H.}~\bibnamefont {Zhong}}, \bibinfo
  {author} {\bibfnamefont {M.~T.}\ \bibnamefont {Batchelor}}, \ and\ \bibinfo
  {author} {\bibfnamefont {C.}~\bibnamefont {Lee}},\ }\href {\doibase
  10.1088/1751-8121/aa5a65} {\bibfield  {journal} {\bibinfo  {journal} {Journal
  of Physics A: Mathematical and Theoretical}\ }\textbf {\bibinfo {volume}
  {50}},\ \bibinfo {pages} {113001} (\bibinfo {year} {2017})}\BibitemShut
  {NoStop}%
\bibitem [{\citenamefont {Fu}(2011)}]{Fu2011}%
  \BibitemOpen
  \bibfield  {author} {\bibinfo {author} {\bibfnamefont {L.}~\bibnamefont
  {Fu}},\ }\href {\doibase 10.1103/PhysRevLett.106.106802} {\bibfield
  {journal} {\bibinfo  {journal} {Phys. Rev. Lett.}\ }\textbf {\bibinfo
  {volume} {106}},\ \bibinfo {pages} {106802} (\bibinfo {year} {2011})},\
  \Eprint {http://arxiv.org/abs/1010.1802} {1010.1802} \BibitemShut {NoStop}%
\bibitem [{\citenamefont {Liu}\ \emph {et~al.}(2013)\citenamefont {Liu},
  \citenamefont {Duan},\ and\ \citenamefont {Fu}}]{Liu2013}%
  \BibitemOpen
  \bibfield  {author} {\bibinfo {author} {\bibfnamefont {J.}~\bibnamefont
  {Liu}}, \bibinfo {author} {\bibfnamefont {W.}~\bibnamefont {Duan}}, \ and\
  \bibinfo {author} {\bibfnamefont {L.}~\bibnamefont {Fu}},\ }\href {\doibase
  10.1103/PhysRevB.88.241303} {\bibfield  {journal} {\bibinfo  {journal} {Phys.
  Rev. B}\ }\textbf {\bibinfo {volume} {88}},\ \bibinfo {pages} {241303}
  (\bibinfo {year} {2013})}\BibitemShut {NoStop}%
\bibitem [{Note1()}]{Note1}%
  \BibitemOpen
  \bibinfo {note} {Ref~\cite {Serbyn2014} observed (but did not discuss) ZQOs
  in this system.}\BibitemShut {Stop}%
\bibitem [{\citenamefont {Hsieh}\ \emph {et~al.}(2012)\citenamefont {Hsieh},
  \citenamefont {Lin}, \citenamefont {Liu}, \citenamefont {Duan}, \citenamefont
  {Bansil},\ and\ \citenamefont {Fu}}]{Hsieh2012}%
  \BibitemOpen
  \bibfield  {author} {\bibinfo {author} {\bibfnamefont {T.~H.}\ \bibnamefont
  {Hsieh}}, \bibinfo {author} {\bibfnamefont {H.}~\bibnamefont {Lin}}, \bibinfo
  {author} {\bibfnamefont {J.}~\bibnamefont {Liu}}, \bibinfo {author}
  {\bibfnamefont {W.}~\bibnamefont {Duan}}, \bibinfo {author} {\bibfnamefont
  {A.}~\bibnamefont {Bansil}}, \ and\ \bibinfo {author} {\bibfnamefont
  {L.}~\bibnamefont {Fu}},\ }\href {\doibase 10.1038/ncomms1969} {\bibfield
  {journal} {\bibinfo  {journal} {Nature Communications}\ }\textbf {\bibinfo
  {volume} {3}},\ \bibinfo {pages} {982} (\bibinfo {year} {2012})}\BibitemShut
  {NoStop}%
\bibitem [{\citenamefont {Okada}\ \emph {et~al.}(2013)\citenamefont {Okada},
  \citenamefont {Serbyn}, \citenamefont {Lin}, \citenamefont {Walkup},
  \citenamefont {Zhou}, \citenamefont {Dhital}, \citenamefont {Neupane},
  \citenamefont {Xu}, \citenamefont {Wang}, \citenamefont {Sankar},
  \citenamefont {Chou}, \citenamefont {Bansil}, \citenamefont {Hasan},
  \citenamefont {Wilson}, \citenamefont {Fu},\ and\ \citenamefont
  {Madhavan}}]{Okada2013}%
  \BibitemOpen
  \bibfield  {author} {\bibinfo {author} {\bibfnamefont {Y.}~\bibnamefont
  {Okada}}, \bibinfo {author} {\bibfnamefont {M.}~\bibnamefont {Serbyn}},
  \bibinfo {author} {\bibfnamefont {H.}~\bibnamefont {Lin}}, \bibinfo {author}
  {\bibfnamefont {D.}~\bibnamefont {Walkup}}, \bibinfo {author} {\bibfnamefont
  {W.}~\bibnamefont {Zhou}}, \bibinfo {author} {\bibfnamefont {C.}~\bibnamefont
  {Dhital}}, \bibinfo {author} {\bibfnamefont {M.}~\bibnamefont {Neupane}},
  \bibinfo {author} {\bibfnamefont {S.}~\bibnamefont {Xu}}, \bibinfo {author}
  {\bibfnamefont {Y.~J.}\ \bibnamefont {Wang}}, \bibinfo {author}
  {\bibfnamefont {R.}~\bibnamefont {Sankar}}, \bibinfo {author} {\bibfnamefont
  {F.}~\bibnamefont {Chou}}, \bibinfo {author} {\bibfnamefont {A.}~\bibnamefont
  {Bansil}}, \bibinfo {author} {\bibfnamefont {M.~Z.}\ \bibnamefont {Hasan}},
  \bibinfo {author} {\bibfnamefont {S.~D.}\ \bibnamefont {Wilson}}, \bibinfo
  {author} {\bibfnamefont {L.}~\bibnamefont {Fu}}, \ and\ \bibinfo {author}
  {\bibfnamefont {V.}~\bibnamefont {Madhavan}},\ }\href {\doibase
  10.1126/science.1239451} {\bibfield  {journal} {\bibinfo  {journal}
  {Science}\ }\textbf {\bibinfo {volume} {341}},\ \bibinfo {pages} {1496}
  (\bibinfo {year} {2013})}\BibitemShut {NoStop}%
\bibitem [{\citenamefont {Wang}\ \emph {et~al.}(2013)\citenamefont {Wang},
  \citenamefont {Tsai}, \citenamefont {Lin}, \citenamefont {Xu}, \citenamefont
  {Neupane}, \citenamefont {Hasan},\ and\ \citenamefont {Bansil}}]{Wang2013}%
  \BibitemOpen
  \bibfield  {author} {\bibinfo {author} {\bibfnamefont {Y.~J.}\ \bibnamefont
  {Wang}}, \bibinfo {author} {\bibfnamefont {W.-F.}\ \bibnamefont {Tsai}},
  \bibinfo {author} {\bibfnamefont {H.}~\bibnamefont {Lin}}, \bibinfo {author}
  {\bibfnamefont {S.-Y.}\ \bibnamefont {Xu}}, \bibinfo {author} {\bibfnamefont
  {M.}~\bibnamefont {Neupane}}, \bibinfo {author} {\bibfnamefont {M.~Z.}\
  \bibnamefont {Hasan}}, \ and\ \bibinfo {author} {\bibfnamefont
  {A.}~\bibnamefont {Bansil}},\ }\href {\doibase 10.1103/PhysRevB.87.235317}
  {\bibfield  {journal} {\bibinfo  {journal} {Phys. Rev. B}\ }\textbf {\bibinfo
  {volume} {87}},\ \bibinfo {pages} {235317} (\bibinfo {year}
  {2013})}\BibitemShut {NoStop}%
\bibitem [{\citenamefont {Assaf}\ \emph {et~al.}(2016)\citenamefont {Assaf},
  \citenamefont {Phuphachong}, \citenamefont {Volobuev}, \citenamefont
  {Inhofer}, \citenamefont {Bauer}, \citenamefont {Springholz}, \citenamefont
  {de~Vaulchier},\ and\ \citenamefont {Guldner}}]{Assaf2016}%
  \BibitemOpen
  \bibfield  {author} {\bibinfo {author} {\bibfnamefont {B.}~\bibnamefont
  {Assaf}}, \bibinfo {author} {\bibfnamefont {T.}~\bibnamefont {Phuphachong}},
  \bibinfo {author} {\bibfnamefont {V.}~\bibnamefont {Volobuev}}, \bibinfo
  {author} {\bibfnamefont {A.}~\bibnamefont {Inhofer}}, \bibinfo {author}
  {\bibfnamefont {G.}~\bibnamefont {Bauer}}, \bibinfo {author} {\bibfnamefont
  {G.}~\bibnamefont {Springholz}}, \bibinfo {author} {\bibfnamefont
  {L.}~\bibnamefont {de~Vaulchier}}, \ and\ \bibinfo {author} {\bibfnamefont
  {Y.}~\bibnamefont {Guldner}},\ }\href {\doibase 10.1038/srep20323} {\bibfield
   {journal} {\bibinfo  {journal} {Scientific Reports}\ }\textbf {\bibinfo
  {volume} {6}},\ \bibinfo {pages} {20323} (\bibinfo {year}
  {2016})}\BibitemShut {NoStop}%
\bibitem [{\citenamefont {Sebastian}\ and\ \citenamefont
  {Proust}(2015)}]{QOCuprates}%
  \BibitemOpen
  \bibfield  {author} {\bibinfo {author} {\bibfnamefont {S.~E.}\ \bibnamefont
  {Sebastian}}\ and\ \bibinfo {author} {\bibfnamefont {C.}~\bibnamefont
  {Proust}},\ }\href {\doibase 10.1146/annurev-conmatphys-030212-184305}
  {\bibfield  {journal} {\bibinfo  {journal} {Annual Review of Condensed Matter
  Physics}\ }\textbf {\bibinfo {volume} {6}},\ \bibinfo {pages} {411} (\bibinfo
  {year} {2015})},\ \Eprint
  {http://arxiv.org/abs/https://doi.org/10.1146/annurev-conmatphys-030212-184305}
  {https://doi.org/10.1146/annurev-conmatphys-030212-184305} \BibitemShut
  {NoStop}%
\bibitem [{\citenamefont {Serbyn}\ and\ \citenamefont {Fu}(2014)}]{Serbyn2014}%
  \BibitemOpen
  \bibfield  {author} {\bibinfo {author} {\bibfnamefont {M.}~\bibnamefont
  {Serbyn}}\ and\ \bibinfo {author} {\bibfnamefont {L.}~\bibnamefont {Fu}},\
  }\href {\doibase 10.1103/PhysRevB.90.035402} {\bibfield  {journal} {\bibinfo
  {journal} {Phys. Rev. B}\ }\textbf {\bibinfo {volume} {90}},\ \bibinfo
  {pages} {035402} (\bibinfo {year} {2014})}\BibitemShut {NoStop}%
\bibitem [{\citenamefont {Chan}\ \emph {et~al.}(2016)\citenamefont {Chan},
  \citenamefont {Chiu}, \citenamefont {Chou},\ and\ \citenamefont
  {Schnyder}}]{Chan2016}%
  \BibitemOpen
  \bibfield  {author} {\bibinfo {author} {\bibfnamefont {Y.-H.}\ \bibnamefont
  {Chan}}, \bibinfo {author} {\bibfnamefont {C.-K.}\ \bibnamefont {Chiu}},
  \bibinfo {author} {\bibfnamefont {M.~Y.}\ \bibnamefont {Chou}}, \ and\
  \bibinfo {author} {\bibfnamefont {A.~P.}\ \bibnamefont {Schnyder}},\ }\href
  {\doibase 10.1103/PhysRevB.93.205132} {\bibfield  {journal} {\bibinfo
  {journal} {Phys. Rev. B}\ }\textbf {\bibinfo {volume} {93}},\ \bibinfo
  {pages} {205132} (\bibinfo {year} {2016})}\BibitemShut {NoStop}%
\bibitem [{\citenamefont {Fang}\ \emph {et~al.}(2016)\citenamefont {Fang},
  \citenamefont {Weng}, \citenamefont {Dai},\ and\ \citenamefont
  {Fang}}]{Fang2016}%
  \BibitemOpen
  \bibfield  {author} {\bibinfo {author} {\bibfnamefont {C.}~\bibnamefont
  {Fang}}, \bibinfo {author} {\bibfnamefont {H.}~\bibnamefont {Weng}}, \bibinfo
  {author} {\bibfnamefont {X.}~\bibnamefont {Dai}}, \ and\ \bibinfo {author}
  {\bibfnamefont {Z.}~\bibnamefont {Fang}},\ }\href {\doibase
  10.1088/1674-1056/25/11/117106} {\bibfield  {journal} {\bibinfo  {journal}
  {Chinese Physics B}\ }\textbf {\bibinfo {volume} {25}},\ \bibinfo {pages}
  {117106} (\bibinfo {year} {2016})}\BibitemShut {NoStop}%
\bibitem [{\citenamefont {Son}\ and\ \citenamefont
  {Spivak}(2013)}]{PhysRevB.88.104412}%
  \BibitemOpen
  \bibfield  {author} {\bibinfo {author} {\bibfnamefont {D.~T.}\ \bibnamefont
  {Son}}\ and\ \bibinfo {author} {\bibfnamefont {B.~Z.}\ \bibnamefont
  {Spivak}},\ }\href {\doibase 10.1103/PhysRevB.88.104412} {\bibfield
  {journal} {\bibinfo  {journal} {Phys. Rev. B}\ }\textbf {\bibinfo {volume}
  {88}},\ \bibinfo {pages} {104412} (\bibinfo {year} {2013})}\BibitemShut
  {NoStop}%
\bibitem [{\citenamefont {Parameswaran}\ \emph {et~al.}(2014)\citenamefont
  {Parameswaran}, \citenamefont {Grover}, \citenamefont {Abanin}, \citenamefont
  {Pesin},\ and\ \citenamefont {Vishwanath}}]{PhysRevX.4.031035}%
  \BibitemOpen
  \bibfield  {author} {\bibinfo {author} {\bibfnamefont {S.~A.}\ \bibnamefont
  {Parameswaran}}, \bibinfo {author} {\bibfnamefont {T.}~\bibnamefont
  {Grover}}, \bibinfo {author} {\bibfnamefont {D.~A.}\ \bibnamefont {Abanin}},
  \bibinfo {author} {\bibfnamefont {D.~A.}\ \bibnamefont {Pesin}}, \ and\
  \bibinfo {author} {\bibfnamefont {A.}~\bibnamefont {Vishwanath}},\ }\href
  {\doibase 10.1103/PhysRevX.4.031035} {\bibfield  {journal} {\bibinfo
  {journal} {Phys. Rev. X}\ }\textbf {\bibinfo {volume} {4}},\ \bibinfo {pages}
  {031035} (\bibinfo {year} {2014})}\BibitemShut {NoStop}%
\bibitem [{\citenamefont {Baum}\ \emph {et~al.}(2015)\citenamefont {Baum},
  \citenamefont {Berg}, \citenamefont {Parameswaran},\ and\ \citenamefont
  {Stern}}]{PhysRevX.5.041046}%
  \BibitemOpen
  \bibfield  {author} {\bibinfo {author} {\bibfnamefont {Y.}~\bibnamefont
  {Baum}}, \bibinfo {author} {\bibfnamefont {E.}~\bibnamefont {Berg}}, \bibinfo
  {author} {\bibfnamefont {S.~A.}\ \bibnamefont {Parameswaran}}, \ and\
  \bibinfo {author} {\bibfnamefont {A.}~\bibnamefont {Stern}},\ }\href
  {\doibase 10.1103/PhysRevX.5.041046} {\bibfield  {journal} {\bibinfo
  {journal} {Phys. Rev. X}\ }\textbf {\bibinfo {volume} {5}},\ \bibinfo {pages}
  {041046} (\bibinfo {year} {2015})}\BibitemShut {NoStop}%
\bibitem [{\citenamefont {Hosur}\ and\ \citenamefont
  {Qi}(2013)}]{HOSUR2013857}%
  \BibitemOpen
  \bibfield  {author} {\bibinfo {author} {\bibfnamefont {P.}~\bibnamefont
  {Hosur}}\ and\ \bibinfo {author} {\bibfnamefont {X.}~\bibnamefont {Qi}},\
  }\href {\doibase https://doi.org/10.1016/j.crhy.2013.10.010} {\bibfield
  {journal} {\bibinfo  {journal} {Comptes Rendus Physique}\ }\textbf {\bibinfo
  {volume} {14}},\ \bibinfo {pages} {857 } (\bibinfo {year} {2013})},\ \bibinfo
  {note} {topological insulators / Isolants topologiques}\BibitemShut {NoStop}%
\bibitem [{\citenamefont {Groth}\ \emph {et~al.}(2014)\citenamefont {Groth},
  \citenamefont {Wimmer}, \citenamefont {Akhmerov},\ and\ \citenamefont
  {Waintal}}]{kwant}%
  \BibitemOpen
  \bibfield  {author} {\bibinfo {author} {\bibfnamefont {C.~W.}\ \bibnamefont
  {Groth}}, \bibinfo {author} {\bibfnamefont {M.}~\bibnamefont {Wimmer}},
  \bibinfo {author} {\bibfnamefont {A.~R.}\ \bibnamefont {Akhmerov}}, \ and\
  \bibinfo {author} {\bibfnamefont {X.}~\bibnamefont {Waintal}},\ }\href
  {\doibase 10.1088/1367-2630/16/6/063065} {\bibfield  {journal} {\bibinfo
  {journal} {New Journal of Physics}\ }\textbf {\bibinfo {volume} {16}},\
  \bibinfo {pages} {063065} (\bibinfo {year} {2014})}\BibitemShut {NoStop}%
\bibitem [{\citenamefont {Devakul}\ \emph {et~al.}()\citenamefont {Devakul},
  \citenamefont {{Y.H. Kwan}}, \citenamefont {{S.L. Sondhi}},\ and\
  \citenamefont {{S.A. Parameswaran}}}]{TDunpub}%
  \BibitemOpen
  \bibfield  {author} {\bibinfo {author} {\bibfnamefont {T.}~\bibnamefont
  {Devakul}}, \bibinfo {author} {\bibnamefont {{Y.H. Kwan}}}, \bibinfo {author}
  {\bibnamefont {{S.L. Sondhi}}}, \ and\ \bibinfo {author} {\bibnamefont {{S.A.
  Parameswaran}}},\ }\href@noop {} {}\bibinfo {note} {{in
  preparation}}\BibitemShut {NoStop}%
\bibitem [{\citenamefont {Rajaraman}(1982)}]{rajaraman1982solitons}%
  \BibitemOpen
  \bibfield  {author} {\bibinfo {author} {\bibfnamefont {R.}~\bibnamefont
  {Rajaraman}},\ }\href {https://books.google.co.uk/books?id=1XucQgAACAAJ}
  {\emph {\bibinfo {title} {Solitons and Instantons}}}\ (\bibinfo  {publisher}
  {North-Holland Publishing Company},\ \bibinfo {year} {1982})\BibitemShut
  {NoStop}%
\bibitem [{\citenamefont {Jain}\ and\ \citenamefont {Kivelson}(1987)}]{JK1}%
  \BibitemOpen
  \bibfield  {author} {\bibinfo {author} {\bibfnamefont {J.~K.}\ \bibnamefont
  {Jain}}\ and\ \bibinfo {author} {\bibfnamefont {S.}~\bibnamefont
  {Kivelson}},\ }\href {\doibase 10.1103/PhysRevA.36.3467} {\bibfield
  {journal} {\bibinfo  {journal} {Phys. Rev. A}\ }\textbf {\bibinfo {volume}
  {36}},\ \bibinfo {pages} {3467} (\bibinfo {year} {1987})}\BibitemShut
  {NoStop}%
\bibitem [{\citenamefont {Jain}\ and\ \citenamefont {Kivelson}(1988)}]{JK2}%
  \BibitemOpen
  \bibfield  {author} {\bibinfo {author} {\bibfnamefont {J.~K.}\ \bibnamefont
  {Jain}}\ and\ \bibinfo {author} {\bibfnamefont {S.}~\bibnamefont
  {Kivelson}},\ }\href {\doibase 10.1103/PhysRevB.37.4111} {\bibfield
  {journal} {\bibinfo  {journal} {Phys. Rev. B}\ }\textbf {\bibinfo {volume}
  {37}},\ \bibinfo {pages} {4111} (\bibinfo {year} {1988})}\BibitemShut
  {NoStop}%
\bibitem [{\citenamefont {Sharpee}\ \emph {et~al.}(2002)\citenamefont
  {Sharpee}, \citenamefont {Dykman},\ and\ \citenamefont {Platzman}}]{Sharpee}%
  \BibitemOpen
  \bibfield  {author} {\bibinfo {author} {\bibfnamefont {T.}~\bibnamefont
  {Sharpee}}, \bibinfo {author} {\bibfnamefont {M.~I.}\ \bibnamefont {Dykman}},
  \ and\ \bibinfo {author} {\bibfnamefont {P.~M.}\ \bibnamefont {Platzman}},\
  }\href {\doibase 10.1103/PhysRevA.65.032122} {\bibfield  {journal} {\bibinfo
  {journal} {Phys. Rev. A}\ }\textbf {\bibinfo {volume} {65}},\ \bibinfo
  {pages} {032122} (\bibinfo {year} {2002})}\BibitemShut {NoStop}%
\bibitem [{\citenamefont {Loss}\ \emph {et~al.}(1992)\citenamefont {Loss},
  \citenamefont {DiVincenzo},\ and\ \citenamefont {Grinstein}}]{SpinTun1}%
  \BibitemOpen
  \bibfield  {author} {\bibinfo {author} {\bibfnamefont {D.}~\bibnamefont
  {Loss}}, \bibinfo {author} {\bibfnamefont {D.~P.}\ \bibnamefont
  {DiVincenzo}}, \ and\ \bibinfo {author} {\bibfnamefont {G.}~\bibnamefont
  {Grinstein}},\ }\href {\doibase 10.1103/PhysRevLett.69.3232} {\bibfield
  {journal} {\bibinfo  {journal} {Phys. Rev. Lett.}\ }\textbf {\bibinfo
  {volume} {69}},\ \bibinfo {pages} {3232} (\bibinfo {year}
  {1992})}\BibitemShut {NoStop}%
\bibitem [{\citenamefont {von Delft}\ and\ \citenamefont
  {Henley}(1992)}]{SpinTun2}%
  \BibitemOpen
  \bibfield  {author} {\bibinfo {author} {\bibfnamefont {J.}~\bibnamefont {von
  Delft}}\ and\ \bibinfo {author} {\bibfnamefont {C.~L.}\ \bibnamefont
  {Henley}},\ }\href {\doibase 10.1103/PhysRevLett.69.3236} {\bibfield
  {journal} {\bibinfo  {journal} {Phys. Rev. Lett.}\ }\textbf {\bibinfo
  {volume} {69}},\ \bibinfo {pages} {3236} (\bibinfo {year}
  {1992})}\BibitemShut {NoStop}%
\bibitem [{\citenamefont {Park}\ and\ \citenamefont {Park}(2000)}]{SpinAB}%
  \BibitemOpen
  \bibfield  {author} {\bibinfo {author} {\bibfnamefont {C.-S.}\ \bibnamefont
  {Park}}\ and\ \bibinfo {author} {\bibfnamefont {D.~K.}\ \bibnamefont
  {Park}},\ }\href {\doibase 10.1142/S0217984900001130} {\bibfield  {journal}
  {\bibinfo  {journal} {Modern Physics Letters B}\ }\textbf {\bibinfo {volume}
  {14}},\ \bibinfo {pages} {919} (\bibinfo {year} {2000})}\BibitemShut
  {NoStop}%
\bibitem [{\citenamefont {Jentschura}\ and\ \citenamefont
  {Zinn-Justin}(2004)}]{JentschuraZinnJustin}%
  \BibitemOpen
  \bibfield  {author} {\bibinfo {author} {\bibfnamefont {U.~D.}\ \bibnamefont
  {Jentschura}}\ and\ \bibinfo {author} {\bibfnamefont {J.}~\bibnamefont
  {Zinn-Justin}},\ }\href {\doibase
  https://doi.org/10.1016/j.physletb.2004.06.077} {\bibfield  {journal}
  {\bibinfo  {journal} {Physics Letters B}\ }\textbf {\bibinfo {volume}
  {596}},\ \bibinfo {pages} {138 } (\bibinfo {year} {2004})}\BibitemShut
  {NoStop}%
\bibitem [{\citenamefont {Dunne}\ and\ \citenamefont
  {\"Unsal}(2014)}]{DunneUnsalWKB}%
  \BibitemOpen
  \bibfield  {author} {\bibinfo {author} {\bibfnamefont {G.~V.}\ \bibnamefont
  {Dunne}}\ and\ \bibinfo {author} {\bibfnamefont {M.}~\bibnamefont
  {\"Unsal}},\ }\href {\doibase 10.1103/PhysRevD.89.105009} {\bibfield
  {journal} {\bibinfo  {journal} {Phys. Rev. D}\ }\textbf {\bibinfo {volume}
  {89}},\ \bibinfo {pages} {105009} (\bibinfo {year} {2014})}\BibitemShut
  {NoStop}%
\bibitem [{\citenamefont {Dorigoni}(2015)}]{dorigoni2015introduction}%
  \BibitemOpen
  \bibfield  {author} {\bibinfo {author} {\bibfnamefont {D.}~\bibnamefont
  {Dorigoni}},\ }\href@noop {} {\enquote {\bibinfo {title} {An introduction to
  resurgence, trans-series and alien calculus},}\ } (\bibinfo {year} {2015}),\
  \Eprint {http://arxiv.org/abs/1411.3585} {arXiv:1411.3585 [hep-th]}
  \BibitemShut {NoStop}%
\bibitem [{\citenamefont {Dunne}\ and\ \citenamefont
  {Ünsal}(2016)}]{DunneUnsal}%
  \BibitemOpen
  \bibfield  {author} {\bibinfo {author} {\bibfnamefont {G.~V.}\ \bibnamefont
  {Dunne}}\ and\ \bibinfo {author} {\bibfnamefont {M.}~\bibnamefont {Ünsal}},\
  }\href {\doibase 10.1146/annurev-nucl-102115-044755} {\bibfield  {journal}
  {\bibinfo  {journal} {Annual Review of Nuclear and Particle Science}\
  }\textbf {\bibinfo {volume} {66}},\ \bibinfo {pages} {245} (\bibinfo {year}
  {2016})}\BibitemShut {NoStop}%
\bibitem [{\citenamefont {Haldane}(1988)}]{HaldaneModel_fromsupp}%
  \BibitemOpen
  \bibfield  {author} {\bibinfo {author} {\bibfnamefont {F.~D.~M.}\
  \bibnamefont {Haldane}},\ }\href {\doibase 10.1103/PhysRevLett.61.2015}
  {\bibfield  {journal} {\bibinfo  {journal} {Phys. Rev. Lett.}\ }\textbf
  {\bibinfo {volume} {61}},\ \bibinfo {pages} {2015} (\bibinfo {year}
  {1988})}\BibitemShut {NoStop}%
\bibitem [{\citenamefont {Kane}\ and\ \citenamefont
  {Mele}(2005)}]{Kane2005_fromsupp}%
  \BibitemOpen
  \bibfield  {author} {\bibinfo {author} {\bibfnamefont {C.~L.}\ \bibnamefont
  {Kane}}\ and\ \bibinfo {author} {\bibfnamefont {E.~J.}\ \bibnamefont
  {Mele}},\ }\href {\doibase 10.1103/PhysRevLett.95.226801} {\bibfield
  {journal} {\bibinfo  {journal} {Phys. Rev. Lett.}\ }\textbf {\bibinfo
  {volume} {95}},\ \bibinfo {pages} {226801} (\bibinfo {year}
  {2005})}\BibitemShut {NoStop}%
\bibitem [{{\relax DLMF}()}]{DLMF_fromsupp}%
  \BibitemOpen
  {\relax DLMF},\ \href {http://dlmf.nist.gov/} {\enquote {\bibinfo {title}
  {{\it NIST Digital Library of Mathematical Functions}},}\ }\bibinfo
  {howpublished} {http://dlmf.nist.gov/, Release 1.0.28 of 2020-09-15},\
  \bibinfo {note} {f.~W.~J. Olver, A.~B. {Olde Daalhuis}, D.~W. Lozier, B.~I.
  Schneider, R.~F. Boisvert, C.~W. Clark, B.~R. Miller, B.~V. Saunders, H.~S.
  Cohl, and M.~A. McClain, eds.}\BibitemShut {Stop}%
\end{thebibliography}%


\begin{thebibliography}{6}%
\makeatletter
\providecommand \@ifxundefined [1]{%
 \@ifx{#1\undefined}
}%
\providecommand \@ifnum [1]{%
 \ifnum #1\expandafter \@firstoftwo
 \else \expandafter \@secondoftwo
 \fi
}%
\providecommand \@ifx [1]{%
 \ifx #1\expandafter \@firstoftwo
 \else \expandafter \@secondoftwo
 \fi
}%
\providecommand \natexlab [1]{#1}%
\providecommand \enquote  [1]{``#1''}%
\providecommand \bibnamefont  [1]{#1}%
\providecommand \bibfnamefont [1]{#1}%
\providecommand \citenamefont [1]{#1}%
\providecommand \href@noop [0]{\@secondoftwo}%
\providecommand \href [0]{\begingroup \@sanitize@url \@href}%
\providecommand \@href[1]{\@@startlink{#1}\@@href}%
\providecommand \@@href[1]{\endgroup#1\@@endlink}%
\providecommand \@sanitize@url [0]{\catcode `\\12\catcode `\$12\catcode
  `\&12\catcode `\#12\catcode `\^12\catcode `\_12\catcode `\%12\relax}%
\providecommand \@@startlink[1]{}%
\providecommand \@@endlink[0]{}%
\providecommand \url  [0]{\begingroup\@sanitize@url \@url }%
\providecommand \@url [1]{\endgroup\@href {#1}{\urlprefix }}%
\providecommand \urlprefix  [0]{URL }%
\providecommand \Eprint [0]{\href }%
\providecommand \doibase [0]{http://dx.doi.org/}%
\providecommand \selectlanguage [0]{\@gobble}%
\providecommand \bibinfo  [0]{\@secondoftwo}%
\providecommand \bibfield  [0]{\@secondoftwo}%
\providecommand \translation [1]{[#1]}%
\providecommand \BibitemOpen [0]{}%
\providecommand \bibitemStop [0]{}%
\providecommand \bibitemNoStop [0]{.\EOS\space}%
\providecommand \EOS [0]{\spacefactor3000\relax}%
\providecommand \BibitemShut  [1]{\csname bibitem#1\endcsname}%
\let\auto@bib@innerbib\@empty
\bibitem [{{\relax DLMF}()}]{DLMF_supp}%
  \BibitemOpen
  {\relax DLMF},\ \href {http://dlmf.nist.gov/} {\enquote {\bibinfo {title}
  {{\it NIST Digital Library of Mathematical Functions}},}\ }\bibinfo
  {howpublished} {http://dlmf.nist.gov/, Release 1.0.28 of 2020-09-15},\
  \bibinfo {note} {f.~W.~J. Olver, A.~B. {Olde Daalhuis}, D.~W. Lozier, B.~I.
  Schneider, R.~F. Boisvert, C.~W. Clark, B.~R. Miller, B.~V. Saunders, H.~S.
  Cohl, and M.~A. McClain, eds.}\BibitemShut {Stop}%
\bibitem [{\citenamefont {Haldane}(1988)}]{HaldaneModel_supp}%
  \BibitemOpen
  \bibfield  {author} {\bibinfo {author} {\bibfnamefont {F.~D.~M.}\
  \bibnamefont {Haldane}},\ }\href {\doibase 10.1103/PhysRevLett.61.2015}
  {\bibfield  {journal} {\bibinfo  {journal} {Phys. Rev. Lett.}\ }\textbf
  {\bibinfo {volume} {61}},\ \bibinfo {pages} {2015} (\bibinfo {year}
  {1988})}\BibitemShut {NoStop}%
\bibitem [{\citenamefont {Kane}\ and\ \citenamefont
  {Mele}(2005)}]{Kane2005_supp}%
  \BibitemOpen
  \bibfield  {author} {\bibinfo {author} {\bibfnamefont {C.~L.}\ \bibnamefont
  {Kane}}\ and\ \bibinfo {author} {\bibfnamefont {E.~J.}\ \bibnamefont
  {Mele}},\ }\href {\doibase 10.1103/PhysRevLett.95.226801} {\bibfield
  {journal} {\bibinfo  {journal} {Phys. Rev. Lett.}\ }\textbf {\bibinfo
  {volume} {95}},\ \bibinfo {pages} {226801} (\bibinfo {year}
  {2005})}\BibitemShut {NoStop}%
\bibitem [{\citenamefont {Zhang}\ \emph {et~al.}(2016)\citenamefont {Zhang},
  \citenamefont {Song},\ and\ \citenamefont {Wang}}]{Zhang2016_supp}%
  \BibitemOpen
  \bibfield  {author} {\bibinfo {author} {\bibfnamefont {L.}~\bibnamefont
  {Zhang}}, \bibinfo {author} {\bibfnamefont {X.-Y.}\ \bibnamefont {Song}}, \
  and\ \bibinfo {author} {\bibfnamefont {F.}~\bibnamefont {Wang}},\ }\href
  {\doibase 10.1103/PhysRevLett.116.046404} {\bibfield  {journal} {\bibinfo
  {journal} {Phys. Rev. Lett.}\ }\textbf {\bibinfo {volume} {116}},\ \bibinfo
  {pages} {046404} (\bibinfo {year} {2016})}\BibitemShut {NoStop}%
\bibitem [{\citenamefont {Liu}\ \emph {et~al.}(2013)\citenamefont {Liu},
  \citenamefont {Duan},\ and\ \citenamefont {Fu}}]{Liu2013_supp}%
  \BibitemOpen
  \bibfield  {author} {\bibinfo {author} {\bibfnamefont {J.}~\bibnamefont
  {Liu}}, \bibinfo {author} {\bibfnamefont {W.}~\bibnamefont {Duan}}, \ and\
  \bibinfo {author} {\bibfnamefont {L.}~\bibnamefont {Fu}},\ }\href {\doibase
  10.1103/PhysRevB.88.241303} {\bibfield  {journal} {\bibinfo  {journal} {Phys.
  Rev. B}\ }\textbf {\bibinfo {volume} {88}},\ \bibinfo {pages} {241303}
  (\bibinfo {year} {2013})}\BibitemShut {NoStop}%
\bibitem [{\citenamefont {Groth}\ \emph {et~al.}(2014)\citenamefont {Groth},
  \citenamefont {Wimmer}, \citenamefont {Akhmerov},\ and\ \citenamefont
  {Waintal}}]{kwant_supp}%
  \BibitemOpen
  \bibfield  {author} {\bibinfo {author} {\bibfnamefont {C.~W.}\ \bibnamefont
  {Groth}}, \bibinfo {author} {\bibfnamefont {M.}~\bibnamefont {Wimmer}},
  \bibinfo {author} {\bibfnamefont {A.~R.}\ \bibnamefont {Akhmerov}}, \ and\
  \bibinfo {author} {\bibfnamefont {X.}~\bibnamefont {Waintal}},\ }\href
  {\doibase 10.1088/1367-2630/16/6/063065} {\bibfield  {journal} {\bibinfo
  {journal} {New Journal of Physics}\ }\textbf {\bibinfo {volume} {16}},\
  \bibinfo {pages} {063065} (\bibinfo {year} {2014})}\BibitemShut {NoStop}%
\end{thebibliography}%
 \end{document}


\title{Supplemental material for `Quantum oscillations  in the zeroth Landau Level and the serpentine Landau fan'}

	\author{T. Devakul }
	\affiliation{Department of Physics, Massachusetts Institute of Technology, Cambridge, Massachusetts 02139, USA}
	\affiliation{Department of Physics, Princeton University, Princeton, New Jersey 08540, USA}
		\author{Yves H. Kwan}
	\affiliation{Rudolf Peierls Centre for Theoretical Physics,  Clarendon Laboratory, Oxford OX1 3PU, UK}
	\author{S. L. Sondhi}
	\affiliation{Department of Physics, Princeton University, Princeton, New Jersey 08540, USA}
		\author{S. A. Parameswaran}
	\affiliation{Rudolf Peierls Centre for Theoretical Physics,  Clarendon Laboratory, Oxford OX1 3PU, UK}

\date{\today }

\begin{abstract}
\end{abstract}

\maketitle
\tableofcontents

\section{Various technical details regarding exact solution}
\subsection{Normalization factor and overlaps}
To derive $\mathcal{N}_n$, 
first write the wavefunction as 
\begin{equation}
\ket{w_n} = W \frac{(a^\dagger)^n}{\sqrt{n!}} \ket{0}
 = W \frac{(a^\dagger)^n}{\sqrt{n!}} W^{-1} W \ket{0}
 = \frac{(A^\dagger_+)^n}{\sqrt{n!}} W \ket{0}
\end{equation}
where 
\begin{equation}
W\ket{0} = e^{\eta (a^\dagger + a)} \ket{0} = e^{\eta^2/2}e^{\eta a^\dagger} e^{\eta a} \ket{0} = e^{\eta^2/2} e^{\eta a^\dagger} \ket{0}
\end{equation}
by the Baker–Campbell–Hausdorff formula.
Expanding
\begin{equation}
e^{\eta a^\dagger} \ket{0} = \sum_{n=0}^{\infty} \frac{\eta^n (a^\dagger)^n}{n!}\ket{0} = \sum_{n=0}^{\infty} \frac{\eta^n}{\sqrt{n!}}\ket{n} = e^{\eta^2/2}\ket{0_-}
\end{equation}
where $\ket{0_-}$ is the normalized coherent state satisfying $a\ket{0_-} = \eta\ket{0_-}$.
Thus, we have
\begin{equation}
\ket{w_n}=e^{\eta^2} \frac{(A_+^\dagger)^n}{\sqrt{n!}}\ket{0_-}
\end{equation}
where
\begin{equation}
\ket{0_-} = \sum_{k=0}^{\infty} e^{-2\eta^2}\frac{(2\eta)^k}{\sqrt{k!}}\ket{k_+}
\end{equation}
such that
\begin{equation}
\begin{split}
\mathcal{N}_n = \braket{w_n|w_n} &= \frac{e^{-2\eta^2}}{n!} \sum_{k=0}^{\infty}\frac{(2\eta)^{2k}}{k!} \bra{k_+}A_+^n (A_+^\dagger)^n \ket{k_+}\\
 &= \frac{e^{-2\eta^2}}{n!} \sum_{k=0}^{\infty}\frac{(2\eta)^{2k}(k+n)!}{(k!)^2}\\
 &= e^{-2\eta^2} {}_1 F_1(1+n;1;4\eta^2)
 \end{split}
\end{equation}
from the definition of the hypergeometric function.

A similar calculation yields, for $n>m$, the overlap
\begin{equation}
\braket{w_n|w_m} = \frac{e^{-2\eta^2}(2\eta)^{n-m}}{(n-m)!}\sqrt{\frac{n!}{m!}} {}_1 F_1(1+n;1+n-m;4\eta^2).
\end{equation}
\subsection{Independence of parity-related states}
In this section, we prove that the state $\ket{\phi} = \sum_{n=0}^N \phi_n \ket{w_n}$ is linearly independent from the state $P_a h \ket{\phi}\equiv \sum_{n=0}^N \bar{\phi}_n \ket{\bar{w}_n}$.

First, notice that each $\ket{w_n}$ can be expanded as a sum of shifted number states $\ket{k_-}$ with $k\leq n$, 
\begin{equation}
\ket{w_n} = e^{\eta^2} \frac{(A_+^\dagger)^n}{\sqrt{n!}}\ket{0_-} = e^{\eta^2} \frac{(A_-^\dagger + 2\eta)^n}{\sqrt{n!}}\ket{0_-} = e^{\eta^2}\sum_{k=0}^{n} \frac{\sqrt{n!}}{\sqrt{k!}(n-k)!}(2\eta)^{n-k} \ket{k_-}
\end{equation}
where $\ket{k_-}$ are orthonormal and satisfies $A_-^\dagger A_-\ket{k_-} = k \ket{k_-}$.
Hence, 
$\ket{\phi} = \sum_{n=0}^N d_n \ket{n_-}$ for a choice of $d_n$.

Next, observe that the states 
\begin{equation}
\ket{F_n} = (A_+^\dagger A_+)^n\ket{0_-} = \left[(A_-^\dagger+2\eta)(A_-+2\eta)\right]^n\ket{0_-}
\end{equation}
also consists of a finite sum of $\ket{k_-}$ with $k\leq n$, as long as $\eta \neq 0$.
The states $\{\ket{F_{n\leq \m}}\}$ are linearly independent since each $\ket{F_n}$ contains a $\ket{n_-}$ component not included in $\{\ket{F_{n^\prime}}\}_{n^\prime<n}$.
Hence, $\{\ket{F_n}\}_{n\leq N}$ spans the same space as $\{\ket{n_-}\}_{n \leq N}$.
We can therefore write
\begin{equation}
\ket{\phi} = \sum_{n=0}^N f_n (A_+^\dagger A_+)^n \ket{0_-}
\end{equation}
for some choice of $f_n$.
The overlap of $\ket{\phi}$ with a number state of the opposite shift, $\ket{k_+}$, is
\begin{equation}
\braket{k_+|\phi} = \sum_{n=0}^N f_n \bra{k_+} (A_+^\dagger A_+)^n \left(e^{-2 \eta^2}\sum_{k^\prime=0}^{\infty} \frac{(2\eta)^{k^\prime}}{\sqrt{k^\prime!}} \ket{k^\prime_+}\right) = \frac{e^{-2\eta^2}(2\eta)^k}{\sqrt{k!}} \sum_{n=0}^N f_n k^n.
\label{eq:polynomial}
\end{equation}

We now prove by contradiction.
Suppose $\ket{\phi}=e^{i\theta} P_a h \ket{\phi}$ are linearly dependent.
Then,
\begin{equation}
\braket{k_+|\phi} = e^{i\theta} \bra{k_+} \left(\sum_{n=0}^N \bar{\phi}_n\ket{\bar{w}_n}\right).
\end{equation}
However, $\ket{\bar{w}_n}$ consists of a finite sum of $\{\ket{k_+}\}_{k\leq n}$, from which it follows that
$\braket{k_+|\phi} = 0$ for all $k > N$.
Eq~\ref{eq:polynomial} in turn implies that $\sum_{n=0}^N f_n k^n=0$ for all $k>\m$.  
However, $\sum_{n\leq \m}f_n k^n=0$ is a (non-trivial) degree-$\m$ polynomial in $k$ which can only have at most $\m$ zeroes.  
This is a contradiction, hence, $\ket{\phi} \neq e^{i\theta} P_a h \ket{\phi}$.

\subsection{Husimi function}
\begin{figure}[t]
\includegraphics{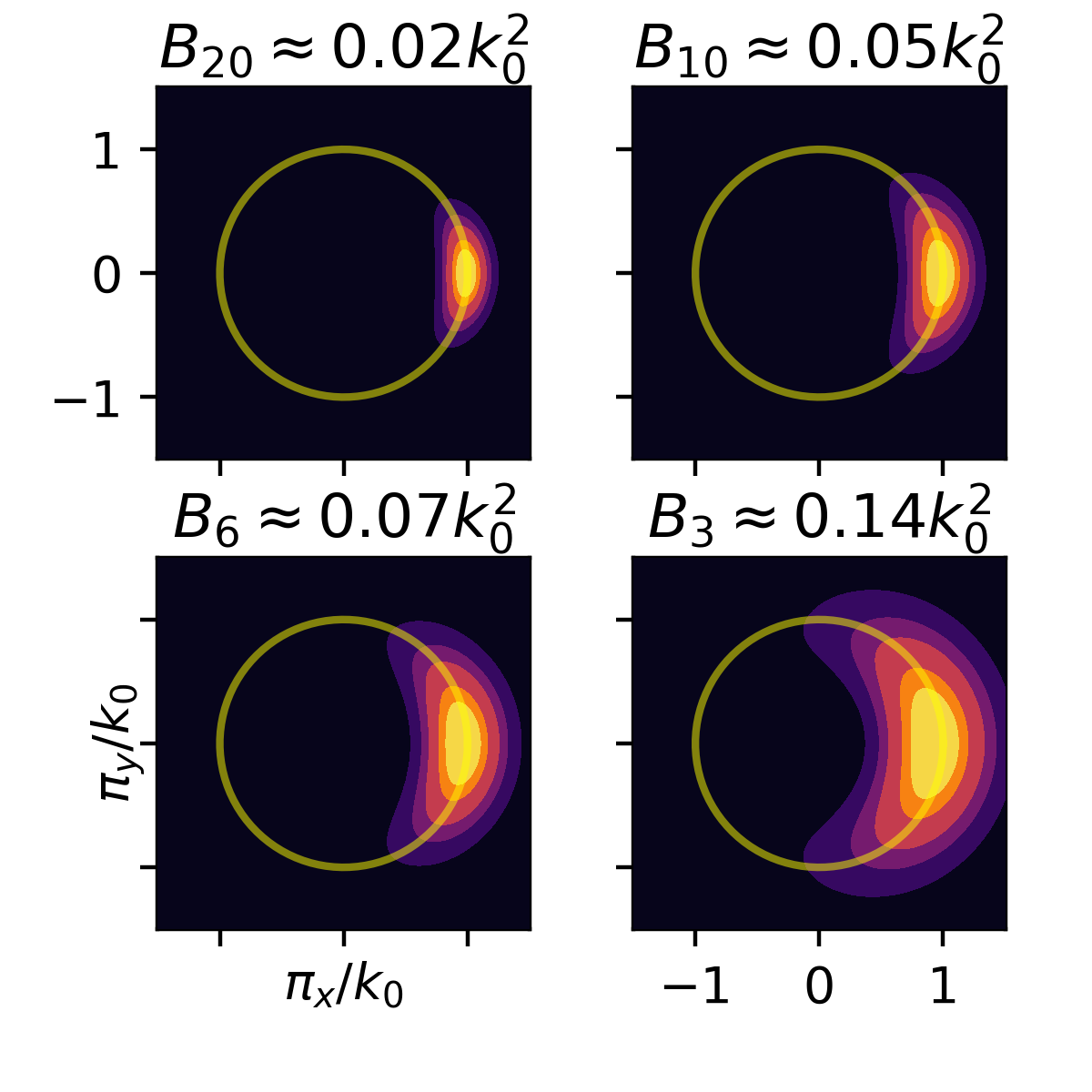}
\caption{A color plot of the Husimi function $Q(\bm{\pi})$ is shown $v=v_c/6$ at a number of magic fields.  
 The yellow circle is the original $v=0$ degenerate ring, which evolves into the low-energy ring along which the central eigenstates spread. }
\label{fig:husimi}
\end{figure}

To visualize the phase space distribution of the zero-energy eigenstates, we compute the Husimi function 
\begin{equation}
Q(\bm{\pi}) = \frac{1}{\pi 2 B_\m\mathcal{N}_\m} |\braket{\bm{\pi}|w_\m}|^2
\end{equation}
which is normalized such that $\int Q(\bm{\pi})d^2\bm{\pi}=1$.
We have
\begin{equation}
Q(\bm{\pi}) 
= \frac{1}{\pi 2 B_\m \mathcal{N}_\m} |\braket{\alpha|W|\m}|^2
\end{equation}
where $a\ket{\alpha}=\alpha\ket{\alpha}$ is a normalized coherent state, $\alpha=(\pi_x+i \pi_y)/\sqrt{2 B_\m}$, and $W=e^{\eta(a+a^\dagger)}$.
Using $W\ket{\alpha} = e^{\eta^2}e^{\eta (\alpha+\alpha^*) } \ket{\alpha+\eta}$,
\begin{equation}
\begin{split}
\braket{\alpha|W|\m} &= e^{\eta^2}e^{\eta(\alpha+\alpha^*)} \braket{\alpha+\eta|\m}\\
&= e^{\eta^2}e^{\eta(\alpha+\alpha^*)} e^{-\frac{|\alpha+\eta|^2}{2}} \frac{(\alpha^*+\eta)^\m}{\sqrt{\m!}}.
\end{split}
\end{equation}
Taking the magnitude squared,
\begin{equation}
\begin{split}
|\braket{\alpha|W|\m}|^2
 &= \frac{1}{\m!}e^{2\eta^2}e^{2\eta(\alpha+\alpha^*)} e^{-|\alpha+\eta|^2} |\alpha+\eta|^{2\m}
 \\
 &= \frac{1}{\m!} e^{2\eta^2}e^{-|\alpha-\eta|^2} |\alpha+\eta|^{2\m}.
 \end{split}.
\end{equation}
Finally, we have
\begin{equation}
Q(\bm{\pi}) = \frac{e^{2\eta^2}|(\pi_x+\frac{v}{v_c} k_0)^2 + \pi_y^2|^\m}{\pi\mathcal{N}_\m(2 B_\m)^{1+\m}  \m!} \exp\left(-\frac{(\pi_x - \frac{v}{v_c} k_0)^2 + \pi_y^2}{2 B_\m}\right) .
\end{equation}
which is shown in Figure~\ref{fig:husimi}.

At large $\m$ (small fields $B_\m\ll B_0$), $Q$ is sharply peaked around the Dirac point $\bm{\pi}=\bm{k}_{0}$, as expected from the low-energy Dirac theory.
As we increase $B$, the inverse magnetic length increases and the states spread in momentum space. Due to the anisotropy of the Dirac cones and the low-energy ``ring'' in the dispersion at $|\bm{\pi}|=k_0$, they eventually form a crescent shape  for $\m>0$.
The state $\ket{\bar{\phi}_0}\propto \ket{\bar{w}_\m}$ spreads similarly  from the other Dirac point.
As $B$ increases, the states spread enough that their  overlap $\braket{\bar{\phi}_0|\phi_0} \sim 
1/\mathcal{N}_N$ is sizable and the two Dirac points can no longer be treated as isolated even at low energy. 
ZQOs and serpentine LL pairing thus reflect departures from the isolated Dirac cone approximation.

\subsection{Computing $\braket{\pi_x}$}\label{sec:pix}
In this section, we compute $\braket{\pi_x}$ of the states $\ket{\phi_0}=\mathcal{N}_{N}^{-\frac{1}{2}}\ket{w_N}$.
First, exactly in terms of hypergeometric functions:
\begin{equation}
\begin{split}
\braket{\pi_x} = \mathcal{N}_N^{-1}\braket{w_N|\pi_x|w_N} &= \mathcal{N}_N^{-1}\sqrt{\frac{B}{2}}\braket{w_N |(a+a^\dagger)|w_N} = \mathcal{N}_N^{-1}\sqrt{\frac{B}{2}}\braket{w_N |(A_-+A_+^\dagger)|w_N} \\
&= \mathcal{N}_N^{-1}\sqrt{\frac{B}{2}} (\sqrt{N} \braket{w_{N}|w_{N-1}} + \sqrt{1+N}\braket{w_N|w_{N+1}})\\
&= \mathcal{N}_N^{-1}\sqrt{\frac{B}{2}}e^{-2\eta^2}(2\eta)\left(N{}_1 F_1(1+N;2;4\eta^2) + (N+1){}_1 F_1(2+N;2;4\eta^2) \right)
\end{split}
\end{equation}

An alternate approach is through the Husimi function.
Let us now compute $\braket{\pi_x}$ to first order in $B_\m$ using the Husimi function, which is relevant to the Haldane and Kane-Mele mass perturbations discussed in Sec~\ref{sec:pert}.

Define $x=\pi_x/\sqrt{2 B_0},y=\pi_y/\sqrt{2 B_0},k=v k_0/(v_c \sqrt{2 B_0})$.
Then,
\begin{equation}
Q(x,y) = C \exp \left\{-(2\m+1)\left[(x-k)^2+y^2\right] + \m \ln\left[(x+k)^2+y^2\right]\right\} 
\equiv C e^{- F(x,y)}
\end{equation}
where $C$ is an unimportant normalization factor.

We wish to approximate $\braket{x} = \int x Q(x,y) / \int Q(x,y) $ to first order in $1/\m$.
The exponent $F(x,y)$ is minimized at $(x_0,y_0) = (\sqrt{k^2 + \m/(2\m+1)},0)$.
Expanding about this point,
\begin{equation}
F(x,y) \approx F_{00} + (x-x_0)^2 F_{20}/2
+ (y-y_0)^2 F_{02}/2
 + (x-x_0)^3 F_{30}/6
 + (x-x_0) (y-y_0)^2 F_{12}/2 + \cdots
\end{equation}
where $F_{ij} = \partial^i_x\partial^j_y F(x,y) |_{x_0,y_0}$.
Then, defining $\tilde{x} = \sqrt{F_{20}}(x-x_0), \tilde{y} = \sqrt{F_{02}}(y-y_0)$, 
\begin{equation}
Q(x,y) \approx C e^{-F_{00}} e^{-\frac{\tilde{x}^2 + \tilde{y}^2}{2}} \left(1 - \frac{F_{30}\tilde{x}^3}{6 \sqrt{F_{20}^3}} - \frac{F_{12}\tilde{x}\tilde{y}^2 }{2\sqrt{F_{20}}F_{02}} + \dots\right),
\end{equation}
and so
\begin{equation}
\begin{split}
\braket{x} &= \frac{
\int d\tilde{x}d\tilde{y} (\frac{\tilde{x}}{\sqrt{F_{20}}}+x_0)e^{-\frac{\tilde{x}^2+\tilde{y}^2}{2}}\left(1 - \frac{F_{30}\tilde{x}^3}{6 \sqrt{F_{20}^3}} - \frac{F_{12}\tilde{x}\tilde{y}^2 }{2\sqrt{F_{20}}F_{02}} + \dots\right)
}{
\int d\tilde{x}d\tilde{y} e^{-\frac{\tilde{x}^2+\tilde{y}^2}{2}}\left(1 - \frac{F_{30}\tilde{x}^3}{6 \sqrt{F_{20}^3}} - \frac{F_{12}\tilde{x}\tilde{y}^2 }{2\sqrt{F_{20}}F_{02}} + \cdots\right)
}\\
&= x_0 + \frac{1}{\sqrt{F_{20}}}
\frac{
\int d\tilde{x}d\tilde{y} e^{-\frac{\tilde{x}^2+\tilde{y}^2}{2}}\left(\tilde{x} - \frac{F_{30}\tilde{x}^4}{6 \sqrt{F_{20}^3}} - \frac{F_{12}\tilde{x}^2\tilde{y}^2 }{2\sqrt{F_{20}}F_{02}} + \cdots\right)
}{
\int d\tilde{x}d\tilde{y} e^{-\frac{\tilde{x}^2+\tilde{y}^2}{2}}\left(1 - \frac{F_{30}\tilde{x}^3}{6 \sqrt{F_{20}^3}} - \frac{F_{12}\tilde{x}\tilde{y}^2 }{2\sqrt{F_{20}}F_{02}} + \cdots\right)
}.
\end{split}
\end{equation}
At large $\m$, $F_{ij}\propto \m$, so the terms in the $\cdots$ are all higher order corrections beyond $1/\m$ and can be neglected.
Evaluating the integrals give
\begin{equation}
\braket{x} \approx x_0 - \frac{1}{2}\left(\frac{F_{30}}{F_{20}^2} + \frac{F_{12}}{F_{20}F_{02}}\right).
\end{equation}
Plugging in for $x_0$, $F_{ij}$, and expanding to first order in $1/\m$ gives
\begin{equation}
\braket{x} \approx \sqrt{k^2+1/2} - \frac{4 k^2 + 1}{16 \m k (2k^2 + 1)}.
\end{equation}
In terms of the original parameters and $B_\m$,
\begin{equation}
\braket{\pi_x} = \sqrt{2 B_0} \braket{x} \approx  k_0\left[1 - \frac{B_N}{ 4 k_0^2}\left(\frac{v}{v_c}+\frac{v_c}{v}\right)\right]
\end{equation}
to first order in $B_\m$.

\subsection{Perturbative estimation of oscillation magnitude}
In this section, we analyze the first order perturbative correction to the energy of the central states near a field $B_\m$.
The Hamiltonian at inverse field $B^{-1}$ is
\begin{equation}
\begin{split}
H(B^{-1}) &= \begin{pmatrix} 
0 & h^\dagger(B^{-1})\\
h(B^{-1}) & 0
\end{pmatrix}\\
h(B^{-1}) &= 
\frac{B v_c}{k_0}\left(a^\dagger a+\frac{1}{2}\right)  + v \frac{\sqrt{B}}{\sqrt{2 }} (a-a^\dagger) - \frac{\Delta}{2} .
\end{split}
\end{equation}
At inverse field $B^{-1}=B_\m^{-1} + \lambda B_0^{-1}$, we have $h(B^{-1}) = h(B_\m^{-1}) + \lambda V + \mathcal{O}(\lambda^2)$, where
\begin{equation}
V 
= -\frac{B_\m}{B_0} \left(\frac{B_\m v_c}{k_0} \left(a^\dagger a + \frac{1}{2}\right) + v\sqrt{\frac{B_\m}{8}}(a-a^\dagger)\right).
\end{equation}
The energy splitting of the two states 
$\ket{\psi_0}=(\mathcal{N}_\m^{-\frac{1}{2}}\ket{w_\m},0), \;\ket{\bar{\psi}_0} = (0,\mathcal{N}_\m^{-\frac{1}{2}}\ket{\bar{w}_\m})$ 
is given by 
\begin{equation}
E_{\mathrm{pert}} = \pm \lambda \braket {\bar{w}_\m | V | w_\m}/\mathcal{N}_\m.
\end{equation}
From the definition of $\ket{w_\m}$, 
\begin{equation}
\begin{split}
\braket {\bar{w}_\m | V | w_\m} = \braket{\m | W^{-1} V W | \m} 
&= 
-\frac{B_\m}{B_0}\braket{\m|\frac{B_\m v_c}{k_0}\left((a^\dagger-\eta)(a+\eta)+\frac{1}{2}\right) + v\sqrt{\frac{B_\m}{8}}(a-a^\dagger+2 \eta)|\m}\\
&= -\frac{B_N}{B_0} \left(\frac{B_\m v_c}{k_0} \left(N  -  \eta^2+ \frac{1}{2}\right)  + v \eta \sqrt{\frac{B_\m}{2}}\right) =  -\frac{B_N v_c}{2 k_0} .
\end{split}
\end{equation}
Therefore, for $B^{-1}$ near $B_\m^{-1}$,
\begin{equation}
E_\mathrm{pert}(B) \approx \pm\frac{(B^{-1}-B_\m^{-1})}{B_0^{-1}} 
\frac{B_\m v_c}{2 k_0 \mathcal{N}_\m}.
\end{equation}
Matching the first derivative in $B^{-1}$ near $B_\m^{-1}$ with the ansatz
\begin{equation}
E_\pm(B) = \pm \mathcal{E}(B) \cos(\pi B_0 / 2 B)
\end{equation}
gives 
\begin{equation}
\mathcal{E}(B_\m) = \frac{B_N v_c}{\pi k_0 \mathcal{N}_N}
\end{equation}
as stated in the main text.
The full expression, after analytic continuation to continuous $B$, is
\begin{equation}
\mathcal{E}(B) = \frac{B v_c }{\pi k_0 
 }\frac{e^{\tilde{B}/B }}{{}_1 F_1\left(\frac{1}{2} + \frac{B_0}{2B};1;2\tilde{B}/B\right)}
\end{equation}
where $\tilde{B} = v^2 k_0^2/v_c^2=(\gamma^2-1)B_0$.

\subsubsection{Limits}
In this section, we find the behavior of the analytically-continued normalization factor, 
\begin{equation}
\mathcal{N} = e^{-\tilde{B}/B}{}_1 F_1\left(\frac{1}{2} + \frac{B_0}{2B};1;2\tilde{B}/B\right) 
\end{equation}
in the limit $B \ll B_0$.
In the limit $B\rightarrow 0$, $B\ll B_0$ and $B\ll \tilde{B}$, which corresponds to both the first and third arguments of the ${}_1 F_1$ becoming large.

Let's first work in the intermediate limit, where $B \ll B_0$ but $B \not \ll \tilde{B}$.
This is a standard limit~\cite[Eq.~13.8.12]{DLMF_supp},
\begin{equation}
{}_1 F_1\left(\frac{1}{2} + \frac{B_0}{2B};1;2\tilde{B}/B\right)  \approx e^{\tilde{B}/B} I_0\left(2\sqrt{\frac{\tilde{B}}{B}\left(1+\frac{B_0}{B}\right)}\right),
\end{equation}
where $I_0$ is a modified Bessel function of the first kind.
The argument of the Bessel function,
\begin{equation}
2\sqrt{\frac{\tilde{B}}{B}\left(1+\frac{B_0}{B}\right)} \approx 2\sqrt{\tilde{B} B_0}/ B 
\end{equation}
at large $B_0/B$.  
Thus,
\begin{equation}
\mathcal{N} \approx  I_0(B^\prime/B_\m)
\end{equation}
where $B^\prime = 2\sqrt{\tilde{B} B_0} = 2 \sqrt{\gamma^2-1}B_0$, 
as stated in the main text.

For the asymptotic small $B$ limit ($B \ll B_0,\tilde{B}$), both arguments of the hypergeometric function diverge.
Define $z\equiv 2\tilde{B}/B$.
 By definition,
\begin{equation}
\mathcal{N}_\m = e^{-z/2}{}_1 F_1(1+N;1;z) = e^{-z/2}\sum_{k=0}^{\infty} \frac{z^k (\m+k)!}{\m! (k!)^2}.
\end{equation}
When $\m$ is large, this will be dominated by terms at large $k$.  
Using Stirling's approximation,
\begin{equation}
\mathcal{N}_\m \approx \frac{e^{-z/2}}{2\pi}\sum_k \sqrt{\frac{\m+k}{\m k^2}} \left(\frac{\m+k}{e}\right)^{\m+k}
\left(\frac{e}{\m}\right)^{\m}\left(\frac{e}{k}\right)^{2k} z^k.
\end{equation}
Reparameterizing $k=\lambda \m$, 
\begin{equation}
\mathcal{N}_\m \approx \frac{e^{-z/2}}{2\pi}\sum_{\lambda}  \sqrt{\frac{1+\lambda}{\m^2 \lambda^2}} \left(\frac{\m(1+\lambda)}{e}\right)^{\m(1+\lambda)}
\left(\frac{e}{\m}\right)^{\m}\left(\frac{e}{\lambda N}\right)^{2\lambda \m} z^{\lambda \m}
\end{equation}
where the sum is over $\lambda = k/\m$ for non-negative integers $k$.
Let us write $z=\alpha(\m+1/2)$, where $\alpha=4 \tilde{B}/B_0 = 4 (\gamma^2-1)$.  
In the large-$\m$ limit, 
\begin{equation}
z^{\lambda \m} = \alpha^{\lambda \m} (\m+1/2)^{\lambda \m} = (\alpha \m)^{\lambda \m} e^{\lambda/2}(1+\mathcal{O}(\m^{-1})).
\end{equation}
Keeping only the leading term, we have 
\begin{equation}
\mathcal{N}_\m \approx \frac{e^{-z/2}}{2\pi}\sum_{\lambda}  \sqrt{\frac{1+\lambda}{\m^2 \lambda^2}} 
e^{\lambda/2}
e^{\m\left[(1+\lambda)\ln(1+\lambda) - 2\lambda \ln\lambda + \lambda\ln\alpha+\lambda\right]}.
\end{equation}
In the limit of large $\m$, the sum can be turned into an integral, $\sum_\lambda\approx \m \int_{-\infty}^{\infty} d\lambda$, and Laplace's method can be applied.  
Define 
\begin{equation}
\phi(\lambda) = (1+\lambda)\ln(1+\lambda) - 2\lambda \ln\lambda + \lambda\ln\alpha+\lambda,
\end{equation}
which is maximized at
\begin{eqnarray}
\phi^\prime(\lambda) = \ln\left(\alpha\frac{1+\lambda}{\lambda^2}\right) = 0\\
\implies \lambda=\lambda_0 \equiv \frac{1}{2}\left(\alpha+\sqrt{\alpha^2+4\alpha}\right).
\end{eqnarray}
At this point, the second derivative is
\begin{equation}
\phi^{\prime\prime}(\lambda_0) = -\frac{2+\lambda_0}{\lambda_0(1+\lambda_0)}.
\end{equation}
Hence, 
\begin{equation}
\begin{split}
\mathcal{N}_\m &\approx \frac{e^{-z/2}}{2\pi}\int d\lambda  \sqrt{\frac{1+\lambda}{ \lambda^2}} 
e^{\lambda/2}
e^{\m\phi(\lambda)}\\
&\approx 
\frac{e^{-z/2}}{2\pi} \sqrt{\frac{1+\lambda_0}{ \lambda_0^2}} 
e^{\lambda_0/2} e^{\m\phi(\lambda_0)}\sqrt{\frac{2\pi}{-\m\phi^{\prime\prime}(\lambda_0)}}\\
&=
\frac{e^{-z/2}e^{\lambda_0/2}}{\sqrt{2\pi \m}}\sqrt{\frac{(1+\lambda_0)^2}{\lambda_0(\lambda_0+2)}} e^{\m\phi(\lambda_0)} \\
&= \frac{1}{\sqrt{2\pi N}}\sqrt{\frac{1+\lambda_0}{\lambda_0(\lambda_0+2)}} e^{\left(N+1/2\right)\left(\lambda_0+\ln(1+\lambda_0)-\alpha/2\right)}
\end{split}
\end{equation}
Finally, some algebra reveals
\begin{equation}
\mathcal{N}_\m = 
\sqrt{\frac{B_\m v_c}{4\pi k_0^2 v}}e^{B^*/B_\m}(1+\mathcal{O}(B_\m/B_0))
\end{equation}
where 
\begin{equation}
B^* = \frac{B_0}{2}\left[\lambda_0-\frac{\alpha}{2}+\ln(1+\lambda_0)\right] = B_0\left[\gamma\sqrt{\gamma^2-1}-\ln\left(\gamma-\sqrt{\gamma^2-1}\right)\right].
\end{equation}
Neglecting the subleading term and going to continuous $B$,
\begin{equation}
\mathcal{N} = \sqrt{\frac{B v_c} {4\pi k_0^2 v}}e^{B^*/B}
\end{equation}
gives rise to the asymptotic form of $\mathcal{E}(B)$ stated in the main text.

\subsection{Various perturbations}\label{sec:pert}
 In this section, we consider how various perturbations to $H$ alter the  ZQOs, focusing on their effect on the central states $\ket{\psi_0}$ and $\ket{\bar{\psi}_0}$ at $B=B_\m$  while working at fixed density corresponding to charge neutrality. 

First, note that parity-preserving perturbations cannot remove LL crossings between opposite-parity states.  This includes
the breaking of particle-hole symmetry by  slightly different effective masses $m_e\neq m_h$ for electron and hole bands, which leads to an overall first-order shift of both $E=0$ states; ZQOs persist since $E_F$ will shift to remain within the central pair at charge neutrality.

A Dirac mass term $m_D\sigma^z$  breaks $\mathcal{P}$ and gaps out the Dirac points, but 
$(H+m_D\sigma^z)^2 = H^2 + m_D^2$ 
has the same eigenstates as $H^2$.
Exactly solvable LLs  at $B_\m$  can once again be found using $h^\dagger h$; the $E_i\neq0$  levels of $H$  remain degenerate but move to $\pm\sqrt{E_i^2+m_D^2}$, preserving their serpentine structure. However $E=0$ states are split by $|2m_D|$, precluding ZQOs, though the modulation of the LL gap can still yield narrow-gap QOs

A Haldane mass $\lambda k_x \sigma^z$ breaks $\mathcal{T}$ and opens a Chern insulator gap $\sim \lambda k_0$ at $B=0$ as it has opposite signs at the two cones~\cite{HaldaneModel_supp}.  
At first order in $\lambda$, it leads to an equal energy shift $E=\lambda\braket{\psi_0|\pi_x|\psi_0}\equiv \lambda\braket{\pi_x}$ for both  central states. $\braket{\pi_x}$ may be computed exactly in terms of hypergeometric functions (see Sec~\ref{sec:pix}): to $O(B^2)$, 
$\braket{\pi_x} \approx  
k_0\left[1 - \frac{B}{ 4 k_0^2}\left(\frac{v}{v_c}+\frac{v_c}{v}\right)\right]$. 
Thus, the shift is largest for $B=0$, and decreases with increasing $B$. 
The central LL pair preserves its serpentine structure but tilts so that the locus of crossing points is no longer parallel to either $E$ or $B$ axes. ZQOs persist if we hold the density fixed, since the $B=0$  Fermi energy is pinned to the conduction/valence band edge (depending on $\text{sign}(\lambda)$).
This is a corollary of the fact~\cite{HaldaneModel_supp} that working at fixed chemical potential requires  increasing/decreasing the density by that of a filled LL, depending on $\text{sign}(\lambda)$. 
Hence it remains within the pair of LLs that evolve into the central pair as $\lambda\to 0$, which thus continue to periodically intersect $E_F$ as $B$ is increased.

Spinful fermions admit a Kane-Mele~\cite{Kane2005_supp} mass term $\lambda s^z k_x \sigma^z$ (where ${s}^\mu$ are spin Pauli matrices) arising from spin-orbit coupling (SOC), and opens a quantum spin Hall (QSH) gap $\sim \lambda k_0$ (note that we have assumed any Zeeman effect is small compared to the orbital effect). 
Since this is a Haldane mass with sign $s^z$, 
for small $\lambda$ the two spin species tilt as above but in opposite directions, to $E_{\uparrow/\downarrow} \approx \pm \lambda \braket{\pi_x}$.
At low fields, the spectrum is fully gapped at neutrality, while at higher fields the oscillation magnitude 
$\mathcal{E}(B)$ can exceed the energy difference $|E_{\uparrow/\downarrow}|$, allowing LLs  to cross $E_F$.
The limit $\lambda=v$ (which in the Rabi analogy corresponds to the Jaynes-Cummings point)
is exactly solvable at all $B$ due to an emergent conservation law. 
As previously noted~\cite{Zhang2016_supp}, ZQOs onset once the gap effectively closes beyond a critical field, but unlike in the gapless case they do not appear for small $B$, although once again the periodic modulation of the gap could yield QOs in thermodynamic quantities.
These results  survive the introduction of a Rashba coupling that breaks the $U(1)$ spin symmetry of the Kane-Mele limit while preserving QSH order, if it preserves $\mathcal{P}$. 
While ZQOs are destroyed if $\mathcal{P}$ is broken, the gap modulation effects persist even in this case. 

\section{TCI surface state solution}
The Hamiltonian for the TCI surface state in a magnetic field is
\begin{equation}
H_{{T}} = v_T( \pi_x s^y - \pi_y s^x) + m_T \tau^x + \delta_T s^x \tau^y  .
\end{equation}
Let us work in the rotated $(\sigma^x,\sigma^z) = (\tau^x,\tau^y)$ Pauli basis. 
Then,
\begin{equation}
H_{{T}} = 
\begin{pmatrix}
0 & - i g_T a^\dagger + \delta_T & m_T & 0 \\
 i g_T a + \delta_T & 0 &  0 & m_T \\
m_T & 0 & 0 & -i g_T a^\dagger - \delta_T\\
0 & m_T & i g_T a- \delta_T & 0
\end{pmatrix},
\end{equation}
where $g_T=v_T \sqrt{2 B}$.
Applying a rotation $a,a^\dagger\rightarrow -i a, i a^\dagger$, and defining $\eta_T = \delta_T/g_T$, $H_{T}$ is reduced to the form
\begin{equation}
H_{{T}} = 
\begin{pmatrix}
0 &  g_T A_+^\dagger  & m_T & 0 \\
 g_T A_+ & 0 &  0 & m_T \\
m_T & 0 & 0 & g_T A_-^\dagger\\
0 & m_T & g_T A_- & 0
\end{pmatrix}
\end{equation}
where $A_{\pm} = a\pm \eta_T$.
Let $\ket{\psi_0} = (0,\ket{\phi^\prime},\ket{\phi},0)$.  
The zero-eigenvalue equations are
\begin{equation}
g_T A_+^\dagger \ket{\phi^\prime} + m_T \ket{\phi} = 0,\;\;m_T \ket{\phi^\prime} + g_T A_- \ket{\phi} = 0.
\end{equation}
Eliminating $\ket{\phi^\prime}$ leads to the equation for $\ket{\phi}$,
\begin{equation}
\left(A_+^\dagger A_- - \frac{m_T^2}{g_T^2}\right) \ket{\phi} = 0,
\end{equation}
which has a solution whenever $\Gamma_T \equiv m_T^2/g_T^2 = \m$ is a positive integer, given by
$\ket{\phi}=  \ket{w_\m} = e^{\eta_T (a+a^\dagger)}\ket{\m}$, where $a^\dagger a \ket{\m} = \m \ket{\m}$.
The zero-energy eigenstate at $B=B_{T,\m}\equiv m_T^2/(2 v_T^2\m)$ is given by
\begin{equation}
\ket{\psi_0} =  \frac{\left(0, -\ket{w_{\m-1}},\ket{w_\m},0\right)}{\sqrt{\mathcal{N}_\m + \mathcal{N}_{\m-1}}} 
\end{equation}
where $\ket{w_{-1}}\equiv 0$, and 
 $\mathcal{N}_n = e^{-2 \eta_T^2} {}_1 F_1(1+n,1, 4 \eta_T^2)$.
Similarly, another zero-energy state exists in the $(\ket{\phi},0,0,\ket{\phi^\prime})$ sector, and is given by
\begin{equation}
\ket{\bar{\psi}_0} = \frac{\left(\ket{\bar{w}_\m},0,0,-\ket{\bar{w}_{\m-1}}\right)}{\sqrt{\mathcal{N}_\m +  \mathcal{N}_{\m-1}}}.
\end{equation}

To determine the magnitude of the energy oscillations, we proceed by a similar procedure as before.
At the inverse field $B^{-1} = B_{T,\m} + \lambda B_{T,1}$, we have
$H_T(B^{-1}) = H_T({B_{T,\m}^{-1}}) + \lambda V$ where
\begin{equation}
V = 
-\frac{v_T B_{T,\m}^{3/2}}{\sqrt{2} B_{T,1}}\begin{pmatrix}
0 &   a^\dagger  & 0 & 0 \\
a & 0 &  0 & 0 \\
0 & 0 & 0 &  a^\dagger\\
0 & 0 &  a & 0
\end{pmatrix}
\end{equation}
The matrix element
\begin{equation}
\braket{\bar{\psi}_0|V|\psi_0} 
= \frac{v_T B_{T,\m}^{3/2}}{\sqrt{2} B_{T,1}}\frac{\braket{\bar{w}_\m|a^\dagger|w_{\m-1}} + \braket{\bar{w}_{\m-1}|a|w_\m}}{\mathcal{N}_\m + \mathcal{N}_{\m-1}}
= \frac{v_T B_{T,\m}^{3/2}}{\sqrt{2} B_{T,1}}\frac{\sqrt{\m}}{\mathcal{N}_\m + \mathcal{N}_{\m-1}} = \frac{B_{T,N} v_T^2}{m_T (\mathcal{N}_N + \mathcal{N}_{N-1})}
\end{equation}
leads to the approximation
\begin{equation}
E_{T,\pm}(B) \approx \mathcal{E}(B) \sin(\pi B_{T,1}/ 2 B)
\label{eq:ETCI}
\end{equation}
with the envelope function given by
\begin{equation}
\mathcal{E}_T(B_{T,\m}) = \frac{2 v_T^2 B_{T,\m}}{\pi m_T (\mathcal{N}_\m + \mathcal{N}_{\m-1})}
\end{equation}
analytically continued to continuous $B$.
Notice the $\pi$ phase shift in Eq~\ref{eq:ETCI} in comparison to the analogous equation for $H$: this can be attributed to Berry curvature of the Dirac point below the Fermi energy at $\bm{k}=0$.
The full form is (including factors of $\hbar$ and $e$)
\begin{equation}
\mathcal{E}_T(B) = \frac{v_T^2 e B}{\hbar \pi m_T}\frac{2e^{\tilde{B}/B}}{{}_1 F_1[1+B_{T,1}/B,1,2\tilde{B}/B] + {}_1 F_1[B_{T,1}/B,1,2\tilde{B}/B]}
\label{eq:Et}
\end{equation}
where $e B_{T,1}/\hbar= m_T^2/(2 v_T^2)$ and $e\tilde{B}/\hbar = \delta_T^2/v_T^2$.
This form is slightly modified from that of $H$, but displays similar qualitative behavior.

The parameters extracted by Ref~\cite{Liu2013_supp} for the surface states of SnTe are
$v_{T,x}=\SI{2.4}{\electronvolt\angstrom}$, $v_{T,y}=\SI{1.3}{\electronvolt\angstrom}$, $m_T=\SI{70}{\milli\electronvolt}$, and $\delta_T=\SI{26}{\milli\electronvolt}$, where $v_{T,x}$ and $v_{T,y}$ accounts for anisotropy of the $v_T$ term (which we have assumed is isotropic).
Rescaling momenta results in an isotropic model with parameter $v_T = \sqrt{v_x v_y} \approx \SI{1.8}{\electronvolt\angstrom}$.
Evaluating Eq~\ref{eq:Et} leads to $\mathcal{E}_T(\SI{15}{\tesla}) \approx \SI{0.1}{\milli\electronvolt} \approx  \SI{1}{\kelvin}$ as quoted in the main text.

\section{Mapping to Rabi model}
The quantum Rabi model  (after rotating $\sigma^{x},\sigma^z\rightarrow\sigma^{z},-\sigma^x$) is given by the Hamiltonian
\begin{equation}
H_R = -\Delta_R \sigma^x + \omega_R a^\dagger a + g_R \sigma^z(a+a^\dagger),
\end{equation}
which gives rise to the Schrodinger equation $(H_R-E_R)\ket{\psi}=0$ in terms of $\ket{\psi}=(\ket{\phi_1},\ket{\phi_2})$,
\begin{equation}
\begin{split}
(\omega_R a^\dagger a + g_R(a+a^\dagger) - E_R)\ket{\phi_1} - \Delta_R \ket{\phi_2} = 0,\\
  (\omega_R a^\dagger a - g_R(a+a^\dagger) - E_R)\ket{\phi_2} - \Delta_R \ket{\phi_1} = 0 .
\end{split}
\end{equation}
The Hamiltonian studied in the main text is
\begin{equation}
H = \omega (a^\dagger a - \delta)\sigma^x - i \omega \eta (a-a^\dagger)  \sigma^y .
\end{equation}
After a gauge transformation $a,a^\dagger\rightarrow i a, -i a^\dagger$,
the corresponding Schrodinger equation is
\begin{equation}
\begin{split}
(\omega a^\dagger a + i\omega \eta (a+a^\dagger) - \omega\delta)\ket{\phi_1} - E \ket{\phi_2} = 0\\
  (\omega a^\dagger a - i\omega \eta (a+a^\dagger) - \omega\delta)\ket{\phi_2} - E \ket{\phi_1} = 0
\end{split}
\end{equation}
which, from inspection, is matched to that of the Rabi model with the mapping
\begin{equation}
\omega_R=\omega,\; g_R=i \omega \eta,\; \Delta_R = E,\; E_R=\omega \delta.
\end{equation}
Note that the mapping involves interchanging energy and model parameters.
This is what allows for the ``analytic continuation'' $g_R = i \omega c$ to imaginary values to be possible, despite both $H$ and $H_R$ being Hermitian.


\section{Tight binding model}
We consider a lattice regularization of $H$ in the form of a tight binding model $H_{TB}$ on a square lattice, which approximately resembles $H$ near the point $\bm{k}=0$ in the Brillouin zone.
Replacing $k_i^2\rightarrow a^{-2}[2-2\cos(k_i a)]$ and $k_i\rightarrow a^{-1}[\sin(k_i a)]$, with $a$ the lattice constant, we have the tight binding model
\begin{equation}
H_{TB} = \left(\frac{1}{2ma^2}[4-2\cos(k_x a)-2\cos(k_y a)] - \frac{\Delta}{2}\right)\tau^z + \frac{v}{a} \sin(k_y a)\tau^y.
\end{equation}
The full second-quantized Hamiltonian is $\mathcal{H}_{TB} = \vec{c}_{\bm{k}}^{\hspace{1mm}\dagger} H_{TB} \vec{c}_{\bm{k}}$,
where $\vec{c}_{\bm{k}}=(c_{\bm{k},1},c_{\bm{k},2})^T$,
$\vec{c}_{\bm{k}}^{\hspace{1mm}\dagger}=(c_{\bm{k},1}^\dagger,c_{\bm{k},2}^\dagger)$, 
$c_{\bm{k},n}$ ($c_{\bm{k},n}^\dagger$) 
is the creation (annihilation) operator for an electron at momentum $\bm{k}$ in orbital index $n=1,2$, and $\tau^\alpha$ are Pauli matrices acting on the orbital index.

Inverse Fourier transforming gives, in real space,
\begin{equation}
\mathcal{H}_{TB} =\sum_{\bm{r}}  
\left(\frac{2}{ma^2}- \frac{\Delta}{2}\right)\vec{c}_{\bm{r}}^{\hspace{1mm}\dagger}
\tau^z \vec{c}_{\bm{r}}
-
\sum_{\bm{r}} \left[\frac{1}{2ma^2}\vec{c}_{\bm{r}+a {\bm{x}}}^{\hspace{1mm}\dagger}
\tau^z \vec{c}_{\bm{r}}
+
\vec{c}_{\bm{r}+a {\bm{y}}}^{\hspace{1mm}\dagger}
\left(\frac{1}{2ma^2}\tau^z + \frac{i v}{2a}\tau^y \right) \vec{c}_{\bm{r}}
 + \mathrm{h.c.}\right]
\end{equation}
a nearest-neighbor tight binding Hamiltonian which has the low energy description of $H$, provided that the Dirac cone positions  $k_xa = \pm \cos^{-1}(1-m\Delta a^2/2) \ll \pi$ are close to $\bm{k}=0$ in the Brillouin zone.

We have numerically verified that $\mathcal{H}_{TB}$ in a magnetic field indeed reproduces the qualitative physics of the continuum model at low energy.
This remains true even when the Dirac cones are located far from the center of the Brillouin zone, as long as they are still connected via the low-energy ring.

\section{Details of numerics for Weyl semimetal}
We consider a discretized version of $H_\mathrm{Weyl}$ from the main text.
The model is defined on a $(L_x,L_y,L_z)$ lattice with open boundary conditions along $y$ and $z$, and periodic boundary conditions.
Using Landau gauge $\bm{A}(\bm{r}) = -B r_y \bm{x}$,  $k_x$ remains a good quantum number and the Hamiltonian at each $k_x$ is
\begin{equation}
\begin{split}
\mathcal{H}(k_x) =& \sum_{\bm{r}} \left\{
\frac{3}{m a^2} - \frac{1}{m a^2}\cos[(k_x + B r_y) a]
-\frac{\Delta}{2}\right\}
\vec{c}_{\bm{r}}^{\hspace{1mm}\dagger} \tau^z \vec{c}_{\bm{r}}
-\sum_{\bm{r}}\left[
\vec{c}_{\bm{r}+a\bm{y}}^{\hspace{1mm}\dagger} 
\left(\frac{1}{2ma^2}\tau^z + \frac{iv}{2a}\tau^y\right)
\vec{c}_{\bm{r}}+\mathrm{h.c.}\right]\\
&-\sum_{\bm{r}}\left[
\vec{c}_{\bm{r}+a\bm{z}}^{\hspace{1mm}\dagger} 
\left(\frac{1}{2ma^2}\tau^z + \frac{iw}{2a}\tau^x\right)
\vec{c}_{\bm{r}}+\mathrm{h.c.}\right] .
\end{split}\label{eq:Hkxweyl}
\end{equation}
Each $k_x a = 2\pi n / L_x$ can then be solved independently and summed over.
For the calculation shown in the main text, we took 
$L_x=200$, $L_y=150$, and $L_z=25,50,100,200,400$.
Sites are labeled according to their position $\bm{r}$, with $r_i/a \in [0,L_i-1]$ where $i=x,y,z$.
All attached leads are also periodic in the $\bm{x}$ direction, and those attached to the top (bottom) extend infinitely in the $(-)\bm{z}$ direction.
Lead $1$ and $2$ span $r_y/a\in[50,70]$ in the $\bm{y}$ direction, and leads $3$ and $4$ span $r_y/a  \in[130,150]$.
The leads are taken to be ideal conductors in the $z$ direction, $H_\mathrm{lead} = -\frac{1}{2 m_\mathrm{lead} a^2}\sum_{\bm{r}} 
\vec{c}_{\bm{r}+a\bm{z}}^{\hspace{1mm}\dagger} 
\sigma^0
\vec{c}_{\bm{r}} $.
The Hamiltonian parameters used are $m=m_\mathrm{lead}=\Delta=1$,$v=1/6$, $w=1/2$, and $a=1/2$ (in appropriate units).
The conductance matrix $C_{ij}$ is computed via the scattering matrix approach using the Kwant code~\cite{kwant_supp}.

Let us first review the zero-field structure of this model.  In the bulk, there are two Weyl points at $\pm\bm{k_0} = \pm(k_0,0,0)$, where $k_0=a^{-1} \cos^{-1}(1-m\Delta a^2/2) \approx 1.01$.
On the $z=0$ and $z=L_z-1$ surface, there is a gapless Fermi arc connecting the two Weyl points whenever $-k_0\leq k_x \leq k_0$.
These Fermi arc surface states are chiral: the top ($z=L_z-1$) surface, they states propagate in the $+\bm{y}$ direction, while on the bottom ($z=0$) surface they propagate in the $-\bm{y}$ direction.

\begin{figure}[t]
\includegraphics[width=0.9\textwidth]{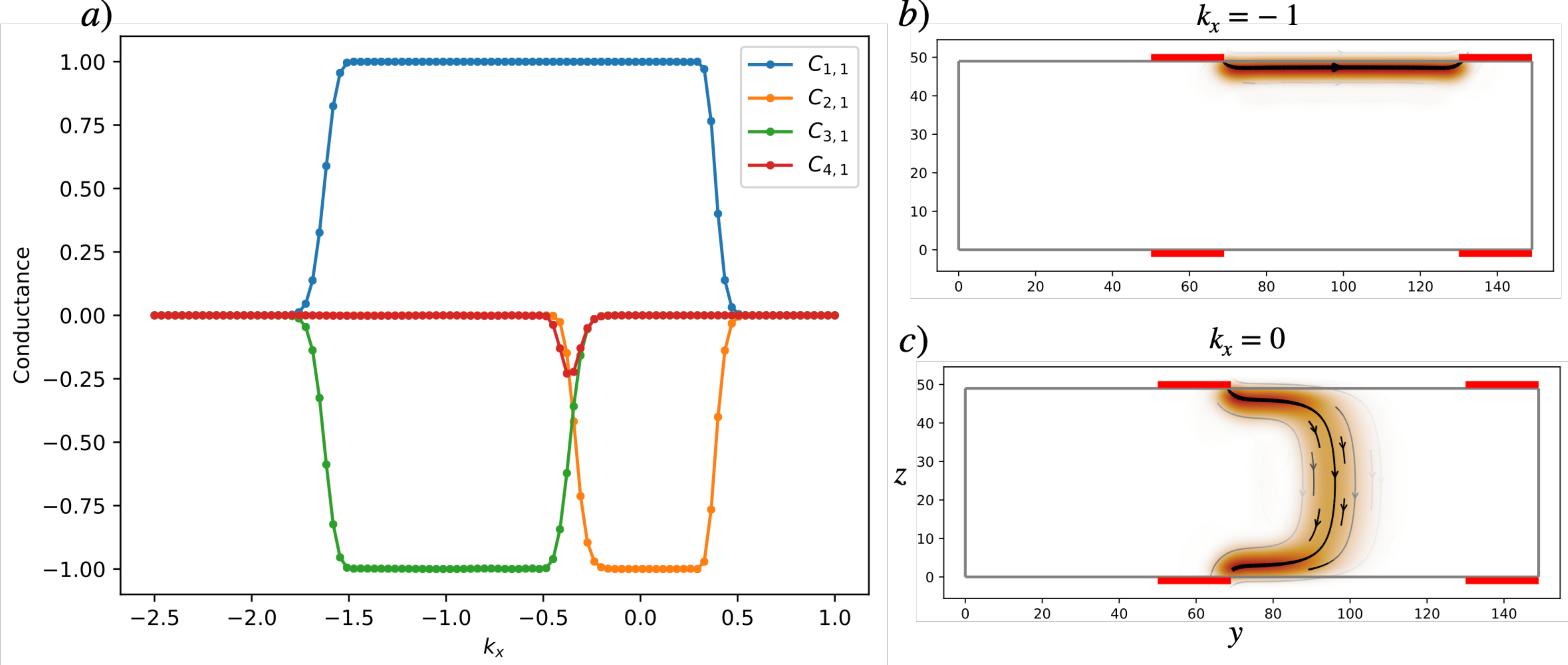}
\caption{$(a)$ Conductance matrix elements (in units of conductance quantum $e^2/h$) as a function of $k_x$ at low field $B=0.02$, for $L_z=50$.
Only the range of $k_x$ for which the conductance is non-zero is shown.
Plots of the current density (discussed in the main text) at $(b)$ $k_x=-1$ and $(c)$ $k_x=0$, showing two qualitatively different current flows from lead $1$.
The gray outline indicates the bounds of the sample, and red rectangles represent the four leads labeled $1,2,3,$ and $4$ going from top-left, bottom-left, top-right, to bottom-right.}
\label{fig:lowBcond}
\end{figure}
\begin{figure}[t]
\includegraphics[width=0.9\textwidth]{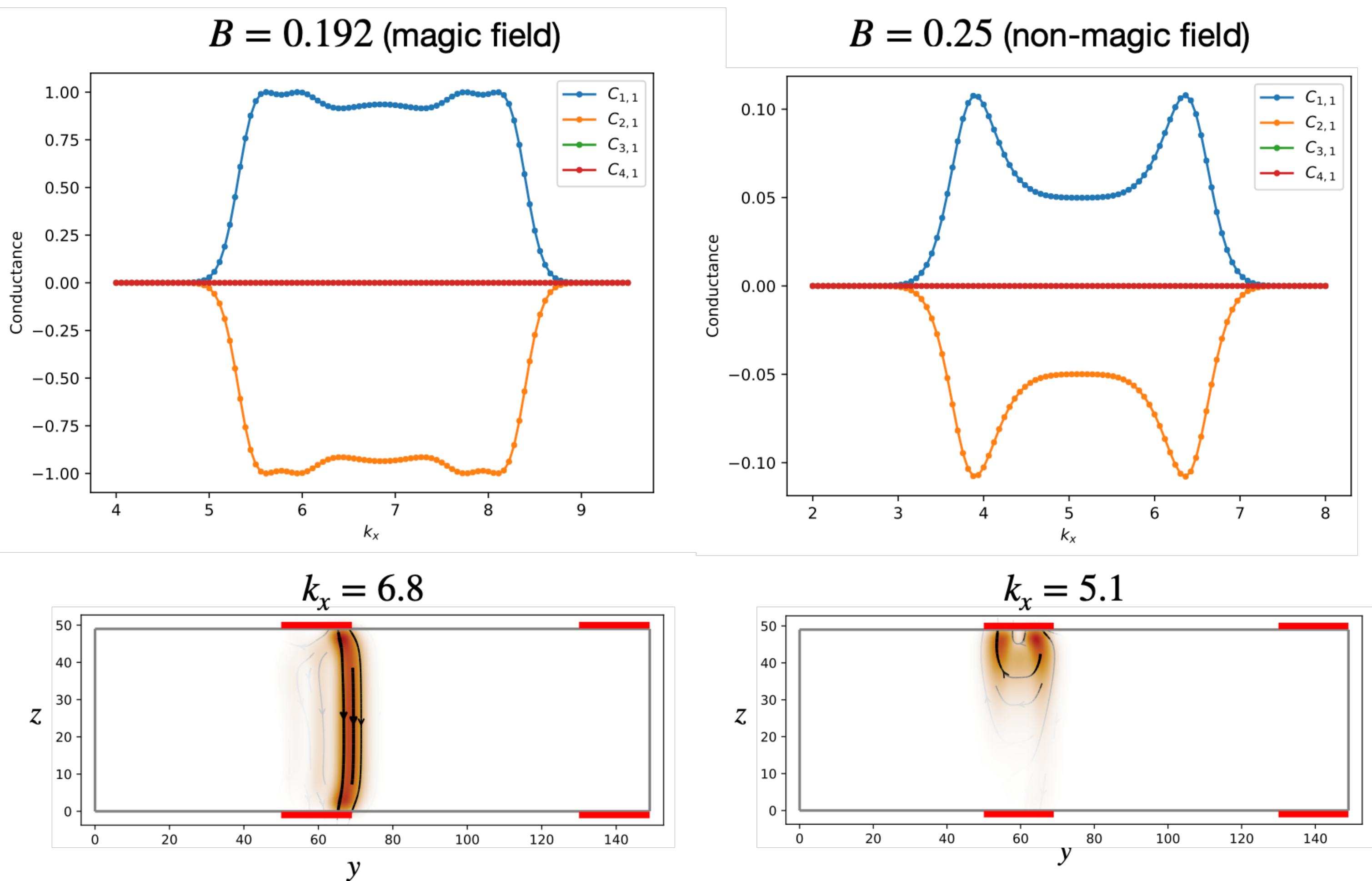}
\caption{Same as Fig~\ref{fig:lowBcond}, but for higher fields at $B=0.192$ (left,approximately a magic field) and $B=0.25$ (right,not a magic field).
The conductance is very small away from the magic field, due to the fact that the bulk ``chiral'' states are not gapless.
}
\label{fig:highBcond}
\end{figure}

To support the explanation for these peaks in the main text, the current carried by states coupled to the lead modes can be directly examined.
Let us first study the low-field behavior at $B=0.02$.
Fig~\ref{fig:lowBcond}a shows some conductance matrix elements as a function of $k_x$, $C_{i 1}(k_x)$, for $L_z=50$, which can be easily understood as follows.
In Landau gauge, $k_x$ is visibly tied to $y$ position, as only the combination $k_x+B r_y$ appears in the Hamiltonian.
Thus, for a fixed $k_x$, assuming that the system can be treated as locally translation invariant, the Fermi arc states only exist within the $r_y$-range $|k_x+B r_y|\leq k_0$.
Thus, lead $1$, which is centered at $y_1=60 a$ (ignoring it's finite width) can only couple to $E=0$ Fermi arc states when $|k_x+B y_1|\leq k_0\implies -1.6\lesssim k_x \lesssim 0.4$.
Similarly, lead $3$ (also on the top edge centered at $y_3=140 a$) can only couple to the Fermi arc states when $|k_x+By_3|\leq k_0 \implies -2.4\lesssim k_x \lesssim -0.4$.
Thus, the Fermi arc states couple leads $1$ and $3$ within the momentum range $-1.6\lesssim k_x \lesssim -0.4$.
Indeed, Fig~\ref{fig:lowBcond}a shows that $C_{11}(k_x)\approx -C_{31}(k_x) = 1$ is precisely the quantum conductance within this range of $k_x$, in agreement with the existence of a single conducting mode (per $k_x$).
In Fig~\ref{fig:lowBcond}b, we show the sum of the current density in all wavefunctions due to the incoming modes from lead $1$, weighted by their net current flow.
It is clear that the current is carried solely by the Fermi arc modes.
On the other hand, within the range $-0.4 \lesssim k_x \lesssim 0.4$, lead $1$ is able to couple to the Fermi arc mode, but lead $3$ is not.
In this regime, electrons propagating along $y$ reach the end of the Fermi arc before reaching lead $3$, and instead are transferred into the bulk chiral states which propagate along $-\bm{z}$.
These traverse entire sample and, upon reaching the bottom surface, are transferred into the Fermi arc surface states propagating along $-\bm{y}$, which is coupled to lead $2$.
Within this range, we therefore have, again, a quantized conductance $C_{11}(k_x)\approx -C_{21}(k_x) = 1$.
The corresponding current density is shown in Fig~\ref{fig:lowBcond}c.
Thus, the main features of Fig~\ref{fig:lowBcond} can all be well understood by this picture.

This picture extends to slightly higher fields as well.  
In Fig~\ref{fig:highBcond}, we show analogous results for $B=0.192$, which is approximately a magic field, and $B=0.25$, which lies between tw o magic fields.  
At such fields, $C_{31}\approx 0$ since no electrons make it to lead $3$.
The conductance $C_{21}\neq 0$ still arises due to the same mechanism as before at magic fields, since the bulk chiral states are perfect (gapless).
Away from a magic field, however, a gap opens up allowing the ``chiral'' states to scatter off each other. 
As a result, the wavefunction becomes exponentially localized on the top edge and, in the limit of large $L_z$, no current makes it to the bottom edge.
This explains the sensitive dependence of $C_{21}$ to ZQOs and the thickness $L_z$, as discussed in the main text.

\section{Non-Lifshitz-Kosevich temperature dependence}

\begin{figure}[t]
\includegraphics{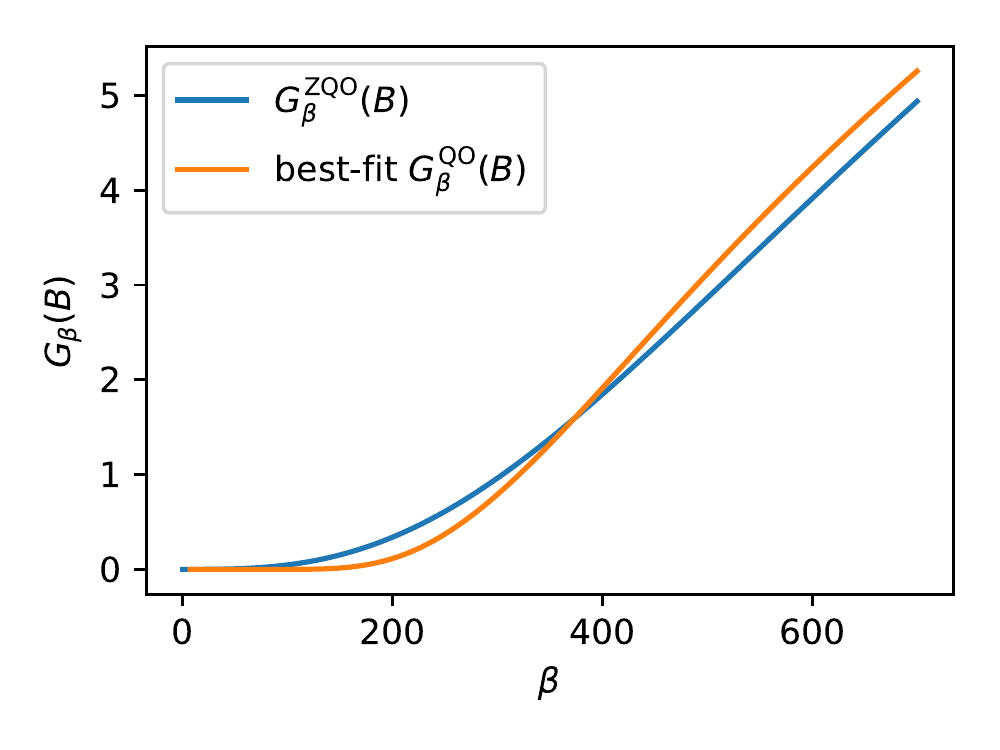}
\caption{The temperature dependence of the low-energy density of states for a model with ZQOs, compared to that of usual QOs which follows the LK form with the best-fit effective mass.}
\label{fig:nonlk}
\end{figure}

In this section, we show that the fact that ZQOs arise due to the motion of just two LLs repeatedly crossing the Fermi energy, rather than the motion of many LLs as in usual QOs, leads to a non-LK temperature dependence.
As can be seen from the serpentine Landau fan (Fig~1a of the main text), there are two energy scales for quantum oscillations at $E=0$: the magnitude of the oscillation (well approximated by $\mathcal{E}(B)$), and the energy gap to the $n=1$ LLs $E_1 \approx \sqrt{2 B v v_c}$.  
We assume we are working at temperatures $T=\beta^{-1} \ll E_1$, such that only the two zeroth LLs are near the Fermi energy.

The low-energy density of states (LEDOS) per unit area is defined
\begin{equation}
 D^{\mathrm{ZQO}}(\beta, B) = -\int d E n_F^\prime(E) \rho_B(E) 
\end{equation}
where $n_F(E)=(1+e^{\beta E})^{-1}$ is the Fermi-Dirac distribution function, $n_F^\prime(E) = \partial_E n_F(E)$, and $\rho_B(E)$ is the density of states per unit area at field $B$, given by $\rho_B(E) = (B/2\pi)\sum_i\delta(E-E_i)$, where $E_i$ are the eigenvalues of the LL problem at magnetic field $B$, and $B/2\pi$ is the LL degeneracy per unit area.
Since $n_F^\prime(E)$ is sharply peaked within an energy range $\beta^{-1}$ of zero, we can safely neglect the contributions of any eigenstates beyond the central two in the limit $\beta^{-1}\ll E_1$.  
Thus,
\begin{equation}
D_\beta(B)\approx -\frac{B}{2\pi}(n_F^\prime(E_+(B)) + n_F^\prime(E_-(B))) = \frac{\beta B}{4\pi \cosh(\beta E_+(B)/2)^2}
\end{equation}
where $E_\pm(B) \approx \pm \mathcal{E}(B) \cos(\pi B_0/ 2 B)$.

In order to define an oscillation magnitude at fixed field, we introduce an additional phase parameter $\phi$, and define
\begin{equation}
E_{\pm}(B,\phi) = \pm \mathcal{E}(B) \cos(\pi B_0/2B + \phi)
\end{equation}
which decouples the oscillating part from the non-oscillating part.
The corresponding LEDOS,
$D^{\mathrm{ZQO}}_\beta(B,\phi)$,
is periodic in $\phi$ and the oscillation magnitude at a fixed $B$ is well defined.
The oscillation magnitude, defined as the $G^{\mathrm{ZQO}}_\beta(B) = \max_\phi D^{\mathrm{ZQO}}_\beta(B,\phi) - \min_\phi D^{\mathrm{ZQO}}_\beta(B,\phi)$, is
\begin{equation}
G^{\mathrm{ZQO}}_\beta(B) = \frac{\beta B}{4\pi}\left[\frac{1}{\cosh(0)^2} - \frac{1}{\cosh(\beta \mathcal{E}(B)/2)^2}\right] = \frac{\beta B}{4\pi} \tanh(\beta\mathcal{E}(B)/2)^2
\label{eq:GZQO}
\end{equation}

Let us now compare with the case of usual QOs arising from a pair of particle and hole Fermi surfaces, $H_{QO}=(\bm{k}^2/2m^* - \mu)\tau^z$, which leads to the usual LL spectrum $E_{\pm n} = \pm\left[(B/m^*)(n+1/2)-\mu\right]$.
Note that unlike with ZQOs, we cannot treat just two eigenstates as isolated.
The LEDOS is
\begin{equation}
D_\beta^{\mathrm{QO}}(B)\approx -\frac{B}{2\pi}\sum_{\pm,n} n_F^\prime(E_{\pm n}(B))) = \frac{\beta B}{4\pi}\sum_{n=0}^{\infty} \frac{1}{\cosh(\beta E_{+ n}(B)/2)^2}.
\end{equation}
Approximating the $n$ sum $\sum_{n=0}^{\infty}\approx\sum_{n=-\infty}^{\infty}$ and applying the Poisson summation formula, this can be rewritten in the form
\begin{equation}
D_\beta^{\mathrm{QO}}(B) = \frac{m^*}{\pi} \sum_{k=-\infty}^{\infty} \frac{k \chi}{\sinh(k\chi)} e^{2\pi i k (\frac{m^*\mu}{B} - \frac{1}{2})} 
\end{equation}
where $\chi=2 \pi^2 m^* / (\beta B)$, which decouples the non-oscillating and oscillating parts.
We can finally obtain the LK form for the oscillation magnitude at fixed $B$,
\begin{equation}
G_\beta^{\mathrm{QO}}(B) = \frac{2m^*}{\pi}\sum_{k=1}^{\infty} [1-(-1)^k]\frac{ k\chi}{\sinh(k\chi)}
\end{equation}
which is to be contrasted with $G_\beta^{\mathrm{ZQO}}(B)$.
Note that there is only a single tunable parameter, the effective mass $m^*$, in $G_\beta^{\mathrm{QO}}(B)$.

As a specific example, consider the case of ZQOs with $v=v_c/6$ as shown in the main text.
For simplicity, we work in units where $m=\Delta=1$.
At a magnetic field $B=0.1$, the two central eigenstates have an oscillation magnitude $\mathcal{E}(B)\approx 0.005$.
The $n=1$ eigenstates are at $E_1\approx 0.15$.  
Thus, Eq~\ref{eq:GZQO} holds within the temperature range $\beta \gg 1/0.15$ where the two central eigenstates are well isolated from the rest of the spectrum.
Fig~\ref{fig:nonlk} shows $G^{\mathrm{ZQO}}_\beta(B)$ for these parameters as a function of $\beta$.  
The non-LK behavior is apparent upon fitting $G_{\beta}^{\mathrm{QO}}(B)$ to $G_{\beta}^{\mathrm{ZQO}}(B)$ within the temperature range $100<\beta < 500$ (well within the range of validity of Eq~\ref{eq:GZQO}) with respect to $m^*$, leading to $m^*\approx 7.1$.
The resulting $G^{\mathrm{QO}}_\beta(B)$, also shown in Fig~\ref{fig:nonlk}, clearly does not well fit $G_{\beta}^{\mathrm{ZQO}}$.

A qualitative difference can be seen in the low-$\beta$ behavior.  
In this limit, the $k=1$ term dominates and $G_\beta^{\mathrm{QO}}(B)\sim \chi/(\pi \sinh[\chi]) \sim \beta^{-1}e^{-2\pi^2 m^*/(\beta B)}$ is exponentially suppressed.
On the other hand, $G_\beta^{\mathrm{ZQO}}(B)\sim \beta^3$ is only polynomial in $\beta$.

The formula shown in the main text, 
$R_{\text{ZLL}}(T) = \frac{\mathcal{E}(B)}{2T} \tanh^2 \frac{\mathcal{E}(B)}{2T}$, is chosen to be dimensionless and match the temperature dependence of Eq~\ref{eq:GZQO}; the factor of $\mathcal{E}(B)$ in front is due to it being the only energy scale in the problem at low temperatures.

\bibliography{supprefs}{}